\documentclass{aa}  

\usepackage{graphicx}
\usepackage{txfonts}
\begin{document}

   \title{J-PLUS: Discovery and characterisation of ultracool dwarfs using Virtual Observatory tools}
   
   \subtitle{II. Second data release and machine learning methodology}
\authorrunning{P.Mas-Buitrago et al.}
\titlerunning{UCDs in J-PLUS DR2}

   \author{P. Mas-Buitrago\inst{\ref{CSIC-INTA},\ref{SVO}}\thanks{E-mail: \tt{pmas@cab.inta-csic.es}},
   E. Solano\inst{\ref{CSIC-INTA},\ref{SVO}},
   A. González-Marcos\inst{\ref{ur}},
   C. Rodrigo\inst{\ref{CSIC-INTA},\ref{SVO}},
   E. L. Martín\inst{\ref{IAC},\ref{ULL},\ref{CSIC}},
   J. A. Caballero\inst{\ref{CSIC-INTA}},
   F. Jiménez-Esteban\inst{\ref{CSIC-INTA},\ref{SVO}},
   P. Cruz\inst{\ref{CSIC-INTA},\ref{SVO}},
   A. Ederoclite\inst{\ref{CEFCA}},
   J.~Ordieres-Mer\'e\inst{\ref{upm}},
   A.~Bello-Garc\'ia\inst{\ref{ovi}},
   R. A. Dupke\inst{\ref{MCTI},\ref{um},\ref{ua}},
   A. J. Cenarro\inst{\ref{cefca2}},
   D. Cristóbal-Hornillos\inst{\ref{cefca2}},
   C. Hernández-Monteagudo\inst{\ref{CEFCA}},
   C. López-Sanjuan\inst{\ref{cefca2}},
   A. Marín-Franch\inst{\ref{cefca2}},
   M. Moles\inst{\ref{CEFCA}},
   J. Varela\inst{\ref{cefca2}},
   H. Vázquez Ramió\inst{\ref{cefca2}},
   J. Alcaniz\inst{\ref{MCTI}},
   L. Sodré Jr.\inst{\ref{iagca}},
   and R. E. Angulo\inst{\ref{DIPC},\ref{IKERBASQUE}}
   }
   \institute{Centro de Astrobiología (CAB), CSIC-INTA, Camino Bajo del Castillo s/n, 28692, Villanueva de la Cañada, Madrid, Spain \label{CSIC-INTA}
        \and
            Spanish Virtual Observatory \label{SVO}
        \and
            Departamento de Ingeniería Mecánica. Universidad de la Rioja. San José de Calasanz 31, 26004 Logroño, La Rioja, Spain \label{ur}
        \and
            Instituto de Astrofísica de Canarias (IAC), Calle  Vía Láctea s/n, 38200 La Laguna, Tenerife, Spain  \label{IAC}
        \and
            Departamento de Astrofísica, Universidad de La Laguna (ULL), 38206 La Laguna, Tenerife, Spain \label{ULL}
        \and
            Consejo Superior de Investigaciones Científicas, 28006 Madrid, Spain \label{CSIC}
        \and
            Centro de Estudios de Física del Cosmos de Aragón (CEFCA), Plaza San Juan 1, 44001 Teruel, Spain \label{CEFCA}
                   \and
            Departamento de Ingenier\'ia de Organizaci\'on, Administraci\'on de Empresas y Estad\'istica, Universidad Polit\'ecnica de Madrid, c/ Jos\'e Guti\'errez Abascal 2, E-28006 Madrid, Spain \label{upm}
        \and 
            Departamento de Construcci\'on e Ingenier\'ia de Fabricaci\'on, Universidad de Oviedo, Pedro Puig Adam, Sede Departamental Oeste, M\'odulo 7, 1$^a$ planta, E-33203 Gij\'on, Spain \label{ovi}
        \and
            Observatório Nacional - MCTI (ON), Rua Gal. José Cristino 77, São Cristóvão, 20921-400 Rio de Janeiro, Brazil \label{MCTI}
        \and
            University of Michigan, Department of Astronomy, 1085 South University Ave., Ann Arbor, MI \label{um}
        \and
            University of Alabama, Department of Physics and Astronomy, Gallalee Hall, Tuscaloosa, AL 35401, USA \label{ua}
        \and
            Centro de Estudios de Física del Cosmos de Aragón (CEFCA), Unidad Asociada al CSIC, Plaza San Juan 1, 44001 Teruel, Spain \label{cefca2}
        \and
            Instituto de Astronomia, Geofísica e Ciências Atmosféricas, Universidade de São Paulo, 05508-090 São Paulo, Brazil \label{iagca}
        \and
            Donostia International Physics Center (DIPC), Paseo Manuel de Lardizabal, 4, 20018, Donostia-San Sebastián, Guipuzkoa, Spain\label{DIPC}
        \and
            IKERBASQUE, Basque Foundation for Science, 48013, Bilbao, Spain \label{IKERBASQUE}
        }

   \date{Received 29 April 2022; Accepted 14 July 2022}

% \abstract{}{}{}{}{} 
% 5 {} token are mandatory
 
  \abstract
   {Ultracool dwarfs (UCDs) comprise the lowest mass members of the stellar population and brown dwarfs, from M7\,V to cooler objects with L, T, and Y spectral types. Most of them have been discovered using wide-field imaging surveys, for which the Virtual Observatory (VO) has proven to be of great utility.}
  {We aim to perform a search for UCDs in the entire Javalambre Photometric Local Universe Survey (J-PLUS) second data release (2\,176\,deg$^2$) following a VO methodology. We also explore the ability to reproduce this search with a purely machine learning (ML)-based methodology that relies solely on J-PLUS photometry.}
   {We followed three different approaches based on parallaxes, proper motions, and colours, respectively, using the \texttt{VOSA} tool to estimate the effective temperatures and complement J-PLUS photometry with other catalogues in the optical and infrared. For the ML methodology, we built a two-step method based on principal component analysis and support vector machine algorithms.}
  {We identified a total of 7\,827 new candidate UCDs, which represents an increase of about 135\,\% in the number of UCDs reported in the sky coverage of the J-PLUS second data release. Among the candidate UCDs, we found 122 possible unresolved binary systems, 78 wide multiple systems, and 48 objects with a high Bayesian probability of belonging to a young association. We also identified four objects with strong excess in the filter corresponding to the Ca~{\sc ii} H and K emission lines and four other objects with excess emission in the H$\alpha$ filter. Follow-up spectroscopic observations of two of them indicate they are normal late-M dwarfs. With the ML approach, we obtained a recall score of 92\,\% and 91\,\% in the  20$\times$20\,deg$^2$ regions used for testing and blind testing, respectively.} 
   {We consolidated the proposed search methodology for UCDs, which will be used in deeper and larger upcoming surveys such as J-PAS and Euclid. We concluded that the ML methodology is more efficient in the sense that it allows for a larger number of true negatives to be discarded prior to analysis with \texttt{VOSA}, although it is more photometrically restrictive.}

\maketitle

%-------------------------------------------------------------------

\section{Introduction} \label{Introduction}

Ultracool dwarfs (UCDs) range from M7\,V to the extended L, T, and Y spectral types, and include very low mass stars, brown dwarfs (BDs), and planetary-mass objects. With effective temperatures of $T_{\rm eff} \lesssim 3\,000$\,K \citep{kirkpatrick2005}, they constitute about 15\,\% of the stellar population in the solar neighbourhood and are abundant throughout the Galaxy \citep{gillon2016, bochanski2010, kirkpatrick2019}. Since UCDs are fainter and smaller than solar-like stars, it is easier to detect transits around them, which gives UCDs a relevant role in the search for Earth-like exoplanets \citep{gillon2017}. As demonstrated in \citet{solano2019}, there is still room to discover nearby UCDs, empowering the discovery and study of new habitable exoplanets in the solar neighbourhood. Furthermore, they play a relevant role in the study of Galactic kinematics and in understanding the boundary between stellar and substellar objects.

In addition to many thousands of late-M dwarfs, the current UCD discoveries include around 2\,000 L and 700 T dwarfs. In tandem, more than 20 Y dwarfs are currently known. In recent years, these discoveries have primarily been driven by wide-field optical and infrared imaging surveys such as the Deep Near Infrared Survey of the Southern Sky \citep[DENIS;][]{denis}, the Sloan Digital Sky Survey \citep[SDSS;][]{sdss}, the Two-Micron All Sky Survey \citep[2MASS;][]{2mass}, the UKIRT Infrared Deep Sky Survey \citep[UKIDSS; ][]{ukidss}, the Wide-Field Infrared Sky Explorer \citep[WISE;][]{wise}, the Visible and Infrared Survey Telescope for Astronomy \citep[VISTA;][]{vista}, and the Panoramic Survey Telescope and Rapid Response System \citep[Pan-STARRS;][]{panstarrs}. The more recent Javalambre Physics of the Accelerated Universe Astrophysical Survey \citep[J-PAS;][]{J-PAS} and the Javalambre Photometric Local Universe Survey \citep[J-PLUS;][]{Cenarro2019}, conceived for the photometric calibration of the former, offer a unique filter system of 56 and 12 filters, respectively. This large number of filters allows for an accurate estimation of stellar parameters such as the effective temperature,  which is crucial for classifying an object as a UCD.

In the years to come, future surveys will produce enormous volumes of data that will compromise traditional processing methods. To overcome this bottleneck, refined machine learning (ML) approaches have gained momentum over the last few years. It is possible that ML will be crucial when it comes to analysing and making predictions as to these volumes of data. This set of techniques has indeed already obtained successful results in several areas of astrophysics. For instance, artificial neural networks (ANNs) have been used to estimate the effective temperature and metallicity of a large subset of stars using broad- and intermediate-band J-PLUS optical photometry \citep{whitten2019}, or to classify galaxies based on their morphology \citep{naim1995}. Nowadays, there is also a variety of ML techniques dedicated to the automated classification of large datasets, as in \citet{wang2021}, where they build a support vector machine \citep[SVM,][]{svm_cortes} to perform a star-galaxy-QSO classification using the 12 J-PLUS optical bands.

The main goal of the study presented here is to extend the search for UCDs initiated in \citet{solano2019} to the J-PLUS second data release (J-PLUS DR2), using a revised Virtual Observatory\footnote{\url{https://ivoa.net/}} (VO) methodology. In addition, we scout two secondary scenarios. The first one is about exploring the capability to reproduce this search with a purely ML-based methodology that relies only on J-PLUS optical photometry. For this purpose, we constructed a two-step ML method based on the principal component analysis (PCA) and the SVM algorithms. The second is to develop an algorithm capable of searching for flares on H$\alpha$ and Ca~{\sc ii} H and K emission lines, again using only the J-PLUS photometry.

This article is organised as follows: In Section \ref{jplus_intro} we describe the J-PLUS survey. Section \ref{Methodology} is dedicated to explaining the methodology we have followed to identify the candidate ultracool dwarfs. In Section \ref{Analysis} we discuss and analyse different properties of our candidates. In Section \ref{known} we compare the candidates found using our methodology with the ultracool dwarfs already found in the J-PLUS field. Sections \ref{ml} and \ref{flares} are devoted to the ML-based approach and the flares search, respectively. Finally, in Section \ref{conclusions} we give a short summary of this work and present our conclusions.

%--------------------------------------------------------------------

\section{J-PLUS} \label{jplus_intro}

J-PLUS is a multi-filter survey conducted from the Observatorio Astrofísico de Javalambre \citep[OAJ;][]{OAJ} in Teruel, Spain. Since it was primarily conceived to ensure the photometric calibration of J-PAS, it uses the second largest telescope at the OAJ, which is the 0.83 m Javalambre Auxiliary Survey Telescope (JAST80). J-PLUS is covering thousands of square degrees of the sky using the panoramic wide-field (2\,deg$^2$ field of view) camera T80Cam \citep{t80cam}, which is equipped with a CCD of 9.2k x 9.2k pixels and a pixel scale of 0.55\,arcsec\,pix$^{-1}$.

While J-PAS will use an unprecedented system of 56 narrow band filters in the optical, the J-PLUS filter system is composed of four broad  (\textit{gSDSS}, \textit{rSDSS}, \textit{iSDSS}, and \textit{zSDSS}), two intermediate (\textit{uJAVA} and \textit{J0861}) and six narrow (\textit{J0378}, \textit{J0395}, \textit{J0410}, \textit{J0430}, \textit{J0515}, and \textit{J0660}) band optical filters. The transmission curves, as well as additional information of these filters, can be found at the Filter Profile Service maintained by the Spanish Virtual Observatory\footnote{\url{http://svo2.cab.inta-csic.es/theory/fps/index.php?&mode=browse&gname=OAJ&gname2=JPLUS}}.

\begin{figure*}
\centering
	\includegraphics[width=\linewidth]{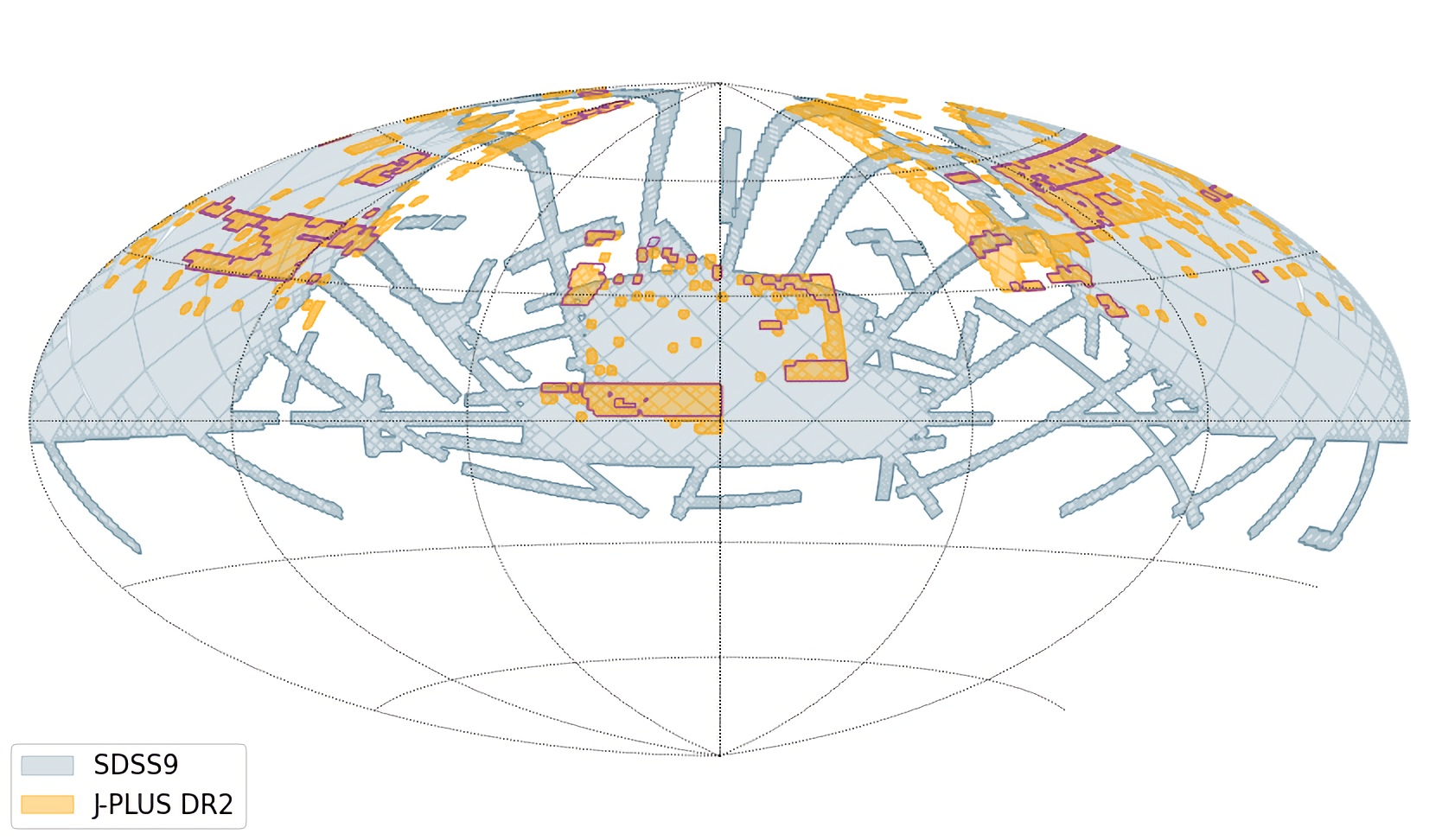}
    \caption{Sky coverage of J-PLUS DR2 (yellow) in $\alpha$, $\delta$ coordinates (centred on $\alpha = 0$ deg, $\delta = 0$ deg, with $\alpha$ rising to the left). The SDSS DR9 footprint is superimposed in blue. The purple line represents the border of the J-PLUS DR1 coverage map.}
    \label{fig:mocs}
\end{figure*}

The J-PLUS DR2, available since November 2020, comprises 1\,088 fields, covering 2\,176\,deg$^2$, observed in all the mentioned optical bands. Fig. \ref{fig:mocs} shows the sky coverage of this release. \citet{jpluscal} presents the updated photometric calibration for the DR2, that was improved by including the metallicity information from LAMOST DR5 in the stellar locus estimation. The limiting magnitudes of the 12 bands can be consulted in the Table 1 of the same paper.

%--------------------------------------------------------------------

\section{Methodology} \label{Methodology}

We divided the sky coverage of J-PLUS DR2 in 37 regions of 20$\times$20\,deg$^2$. To cope with the fact that queries to the J-PLUS archive are limited to 1 million objects, we decided to tessellate each region into smaller circular subregions of 1 deg radius. We made use of \texttt{TOPCAT}\footnote{\url{http://www.star.bris.ac.uk/~mbt/topcat/}} \citep{Taylor2005} to cross-match each tessellated region with the J-PLUS DR2 sky coverage in order to avoid searching regions of the sky that are not covered by it.

We used the package \texttt{STILTS}\footnote{\url{http://www.star.bris.ac.uk/~mbt/stilts/}} \citep{Taylor2006} to query the J-PLUS DR2 database through the Virtual Observatory TAP protocol. This allowed us to write ADQL\footnote{\url{https://www.ivoa.net/documents/REC/ADQL/ADQL-20081030.pdf}} code to search over all 20$\times$20\,deg$^2$ regions iteratively. A typical ADQL query example looks like this:

\begin{verbatim}
SELECT objs.filter_id,objs.alpha_j2000,
    objs.delta_j2000,objs.class_star,
    objs.mag_aper_6_0,objs.mag_err_aper_6_0,
    objs.mask_flags,imgs.aper_cor_6_0,
    imgs.aper_cor_err_6_0 
FROM jplus.MagABSingleObj as objs,
    jplus.TileImage as imgs 
WHERE objs.tile_id = imgs.tile_id
AND objs.alpha_j2000 between 2 and 5 
AND objs.delta_j2000 between 2 and 3 
AND objs.flags=0 
AND objs.filter_id between 1 and 4 
AND objs.class_star>0.1
\end{verbatim}

In our case, we used the 6 arcsec diameter aperture photometry, since the aperture correction to pass 6 arcsec aperture magnitudes to total magnitudes for point-like sources is available in the J-PLUS DR2 database. We constrained the search to records with good photometric conditions by imposing  \texttt{flags=0} (no \texttt{SExtractor} flags\footnote{\url{https://sextractor.readthedocs.io/en/latest/Flagging.html}}). Since object detection is performed independently on each filter, this means that for each source the \texttt{flags=0} condition is applied at the filter level. We also required \texttt{class$\_$star > 0.1}. We were not very restrictive with \texttt{class$\_$star} (\texttt{SExtractor} stellarity index) in order not to loose faint sources that may appear as extended objects.

For each 20$\times$20\,deg$^2$ region, we concatenated the data for the corresponding circular subregions into a single table and removed duplicated instances (tessellated areas may overlap). As UCDs emit most of their flux at longer wavelengths, for the methodology described in Sects. \ref{parallax}, \ref{pm} and \ref{colordiagram}, we only considered the relevant filters for these objects, i.e., the reddest ones (filter IDs 1$-$4 and 10$-$12 in the J-PLUS DR2 database, see Table \ref{tab:filters}). Even so, we stored the data for all filters separately, as we required them for the flare detection workflow described in Sect. \ref{flares}. Finally, we used the CDS \texttt{X-Match} service\footnote{\url{http://cdsxmatch.u-strasbg.fr/}} in \texttt{TOPCAT} with \textit{Gaia} EDR3 J2016 (reference epoch 2016.0), using a 3 arcsec radius, to obtain the astrometric information. In those cases where more than one counterpart exists in the search region, only the nearest one was considered. In Sects. \ref{parallax}, \ref{pm} and \ref{colordiagram} we describe the analysis carried out for each 20$\times$20\,deg$^2$ region separately.

\begin{table}
 \caption{J-PLUS filter information, taken from the J-PLUS DR2 database, sorted from shortest to longest wavelength.}
 \centering
 \label{tab:filters}
 \begin{tabular}{l c c}
 
  \hline\hline
  \noalign{\smallskip}

  Filter ID & Filter \\
  
  \noalign{\smallskip}
  \hline
  \noalign{\smallskip}

  5.0 & $u$ \\
  6.0 & $J0378$ \\
  7.0 & $J0395$ \\
  8.0 & $J0410$ \\
  9.0 & $J0430$ \\
  2.0\,$^{a}$ & $g$ \\
  10.0\,$^{a}$ & $J0515$ \\
  1.0\,$^{a}$ & $r$ \\
  11.0\,$^{a}$ & $J0660$ \\
  3.0\,$^{a}$ & $i$ \\
  12.0\,$^{a}$ & $J0861$ \\
  4.0\,$^{a}$ & $z$ \\
  
  \noalign{\smallskip}
  \hline
 \end{tabular}
 \tablefoot{$^{(a)}$ Relevant filters (reddest ones) for UCDs search.}
\end{table}

\subsection{Parallax-based selection} \label{parallax}

\begin{figure*}
    \centering
    	\includegraphics[width=8.5cm]{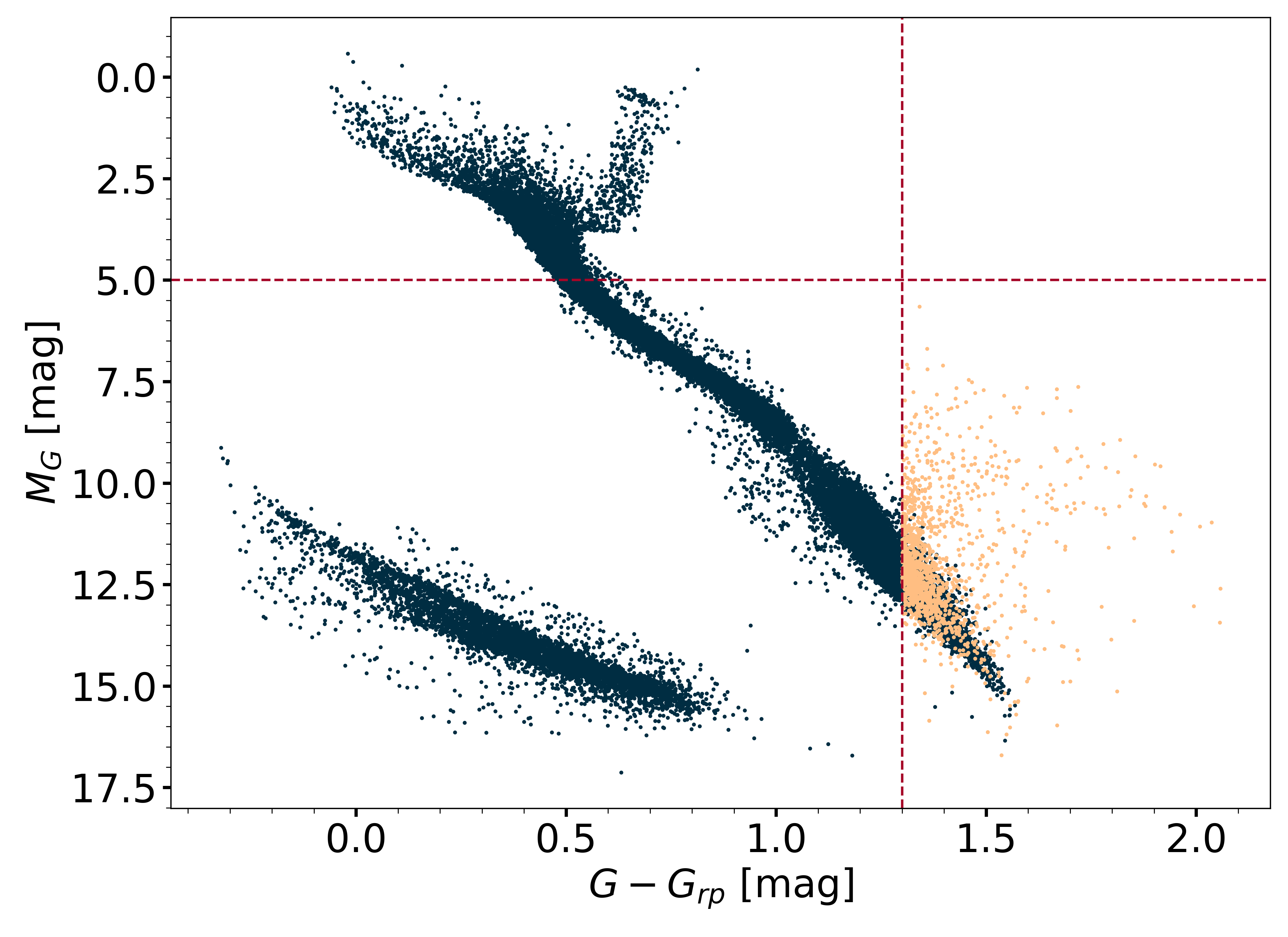}
    	\qquad
    	\includegraphics[width=8.5cm]{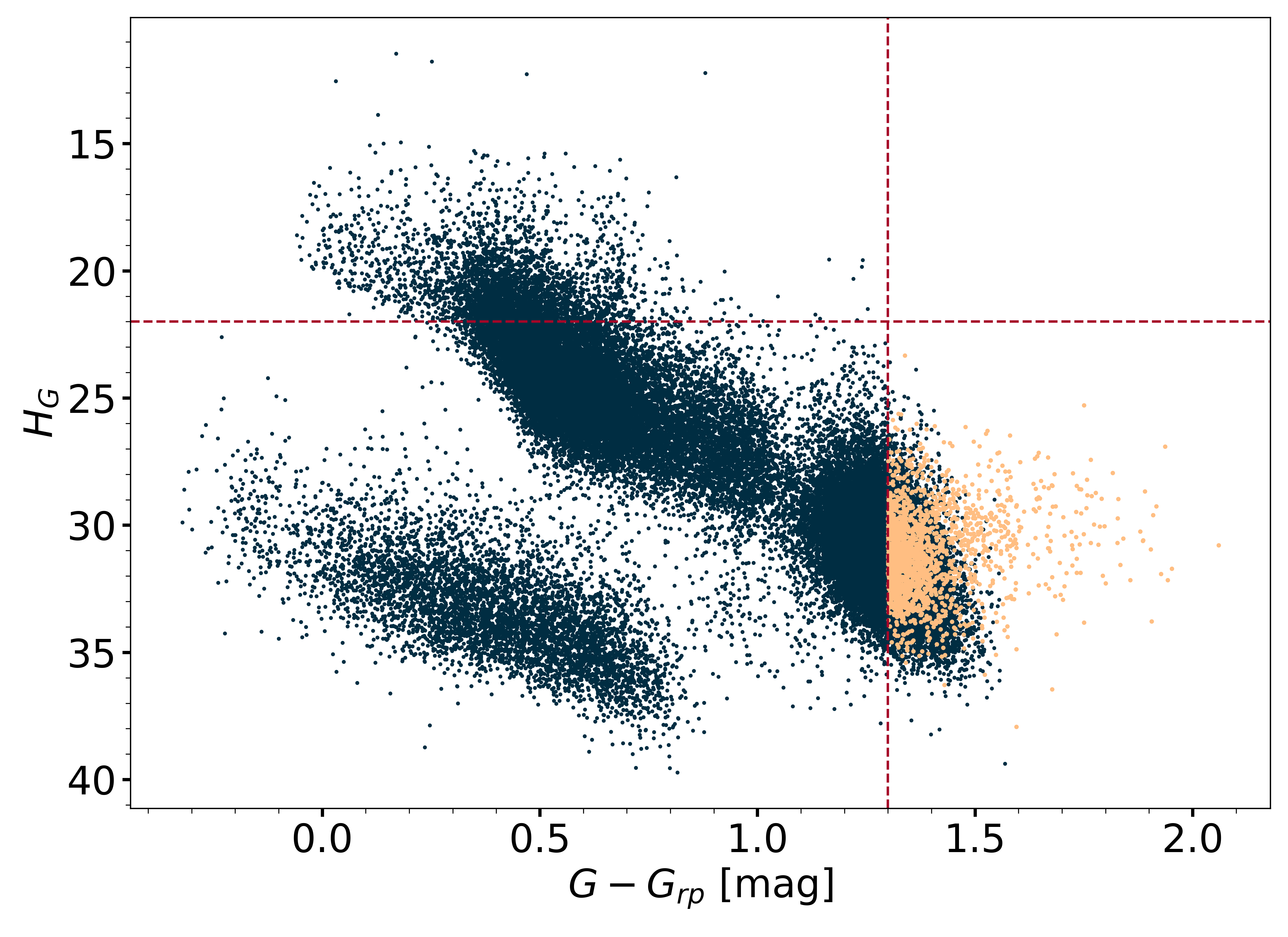}
    \caption{Location of the objects shortlisted as candidate UCDs via astrometric selection, for an example 20$\times$20\,deg$^2$ region, in a colour-magnitude (left) and a reduced proper motion (right) diagram built using \textit{Gaia} EDR3 sources with parallaxes larger tan 10 mas (dark blue dots). The vertical and horizontal red lines mark the boundaries for a source to be shortlisted as candidate UCD. Sources fulfilling these conditions are overplotted in yellow.}
    \label{fig:astro_diagrams}
\end{figure*}

From the cross-matched sample, we only kept sources with relative errors of less than 20\,\% in parallax and less than 10\,\% in both $G$ and $G_{\rm RP}$ photometry. With these objects, we constructed a colour-magnitude diagram (see the left panel of Fig. \ref{fig:astro_diagrams}), where the absolute \textit{Gaia} magnitude in the $G$ band was estimated using

\begin{equation}
    M_G=G + 5\log{\varpi} + 5,
	\label{eq:absoluteg}
\end{equation}

\noindent where $G$ is the \textit{Gaia} apparent magnitude and $\varpi$ is the parallax in arcseconds. To obtain a shortlist of candidate UCDs, we adopted a colour cut of $G - G_{\rm RP} > 1.3$\,mag, which corresponds to spectral types M5\,V or later according to the updated version of Table 5 in \citet{pecaut2013} \footnote{\label{mamajek}\url{http://www.pas.rochester.edu/~emamajek/EEM_dwarf_UBVIJHK_colors_Teff.txt}}, and an absolute magnitude limit of $M_G > 5$\,mag to leave aside the red giant branch.

\subsection{Proper motion-based selection} \label{pm}

Ultracool dwarfs may have photometric and morphological properties similar to those of objects such as giants, quasi-stellar objects (QSOs) or distant luminous red galaxies \citep[e.g.][]{Caballero2018, theissen2016, theyssen2017}. Assuming nearby objects will have high proper motions, reduced proper motion diagrams are a reliable tool for discriminating between nearby stellar populations and distant sources.

From the cross-matched sample introduced in Sect. \ref{Methodology}, we only kept sources with a relative error of less than 20\,\% in both proper motion components and less than 10\,\% in both $G$ and $G_{\rm RP}$ photometry. Furthermore, we only took into account sources with non-zero proper motion, i.e., sources with, at least, one of the proper motion components greater (in absolute value) than three times the associated error.

The right panel of Fig. \ref{fig:astro_diagrams} shows the reduced proper motion diagram defined as:

\begin{equation}
    H_G=G + 5\log{\mu} + 5,
	\label{eq:reducedpm}
\end{equation}

\noindent where $G$ is the \textit{Gaia} apparent magnitude and $\mu$ is the total proper motion in mas\,yr$^{-1}$. Of these sources, we filtered out those already pre-selected in the parallax-guided analysis described in Sect. \ref{parallax} and shortlisted as candidate UCDs those fulfilling the condition $G - G_{\rm RP} > 1.3$\,mag, and with a reduced proper motion $H_G > 22$\,mag to leave aside the red giant branch.

As discussed in Sect. \ref{Methodology}, the cross-match with \textit{Gaia} EDR3 J2016 is done using a 3 arcsec radius. Since J-PLUS DR2 is based on images collected from November 2015 to February 2020, we might miss some objects with a proper motion larger than 750\,mas\,yr$^{-1}$, as they could fall outside this 3\,arcsec radius. However, we decided not to increase the radius to avoid finding erroneous counterparts.

\subsection{Photometry-based selection} \label{colordiagram}

In the first two criteria (colour-magnitude and reduced proper motion diagrams) we are imposing parallax and proper motion constraints respectively, which makes these methods dependent on \textit{Gaia} astrometric information. This means that objects with good photometry but poor astrometry will be excluded from the lists of candidate UCDs. To solve this limitation, in this section we describe a method solely dependent on photometric information. This procedure consisted of two separate steps. First, we built a colour-colour diagram with the purpose of defining a colour cut to identify the UCD locus. Then, we applied this criterion to each 20$\times$20\,deg$^2$ region independently to obtain a shortlist of candidate UCDs.

To built the colour-colour diagram, we first searched in J-PLUS DR2 for true extended sources, defined as sources having \texttt{class$\_$star < 0.01}. Likewise, true point sources were defined as sources with \texttt{class$\_$star > 0.99}. Then, we performed a cross-match with 2MASS and built a $J - K_s$ (2MASS) vs. $r-z$ colour-colour diagram to separate the two types of sources.  As discussed in Sect. \ref{pm}, QSOs may have morphometric properties similar to those of UCDs, so it is crucial to also discriminate between these two types in the colour-colour diagram.

\begin{figure}
	\includegraphics[width=\columnwidth]{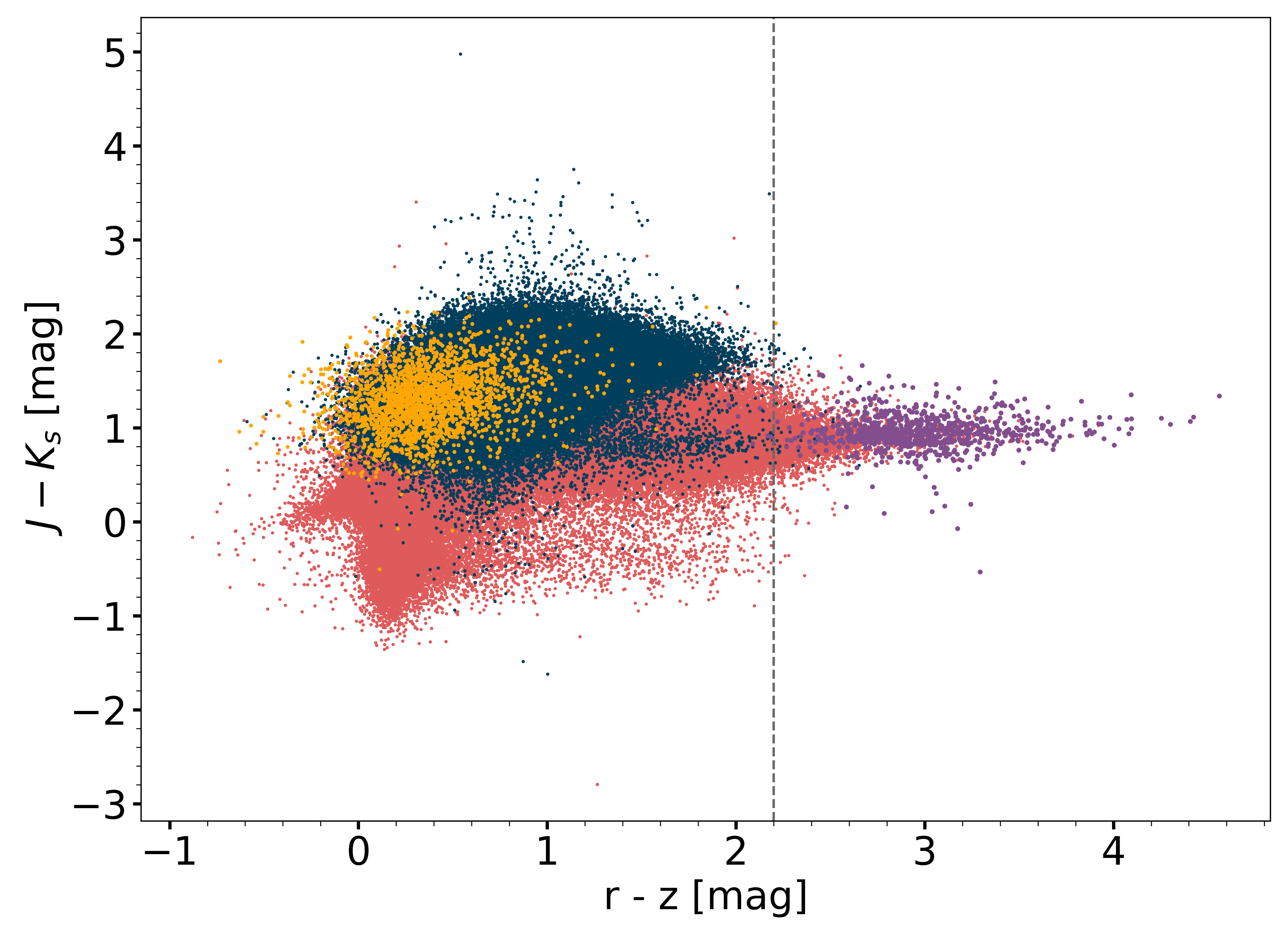}
    \caption{Colour-colour diagram built using true extended (dark blue) and true point (red) sources. Yellow dots represent the sample of 10\,481 QSOs. Purple dots represent the shortlisted candidate UCDs obtained by parallax-guided and proper motion-guided methods. The vertical grey line marks the $r-z > 2.2$\,mag limit for a source to be shortlisted as candidate UCD.}
    \label{fig:colour}
\end{figure}

Fig. \ref{fig:colour} shows the different types of objects in a colour-colour diagram. For the sample of QSOs, we cross-matched the SDSS-DR12 Quasar Catalog\footnote{\url{http://cdsarc.u-strasbg.fr/viz-bin/cat/VII/279}} with the J-PLUS DR2. To define the UCD locus, we overplotted in this diagram the candidate UCDs obtained by the methods described in Sects. \ref{parallax} and \ref{pm} for the region $\alpha$: 0 -- 20\,deg; $\delta$: 0 -- 20\,deg. As a compromise to balance the extended object contamination and the loss of candidate UCDs, we defined the UCD locus as the region fulfilling $r-z > 2.2$\,mag and applied this criterion to all the sources of each 20x20 deg$^2$ region. Of the sources fulfilling it, we filtered out those already pre-selected in the analysis described in Sects. \ref{parallax} and \ref{pm} and shortlisted the remaining ones as candidate UCDs.

\subsection{VOSA filtering} \label{vosa}

\begin{figure}
	\includegraphics[width=\columnwidth]{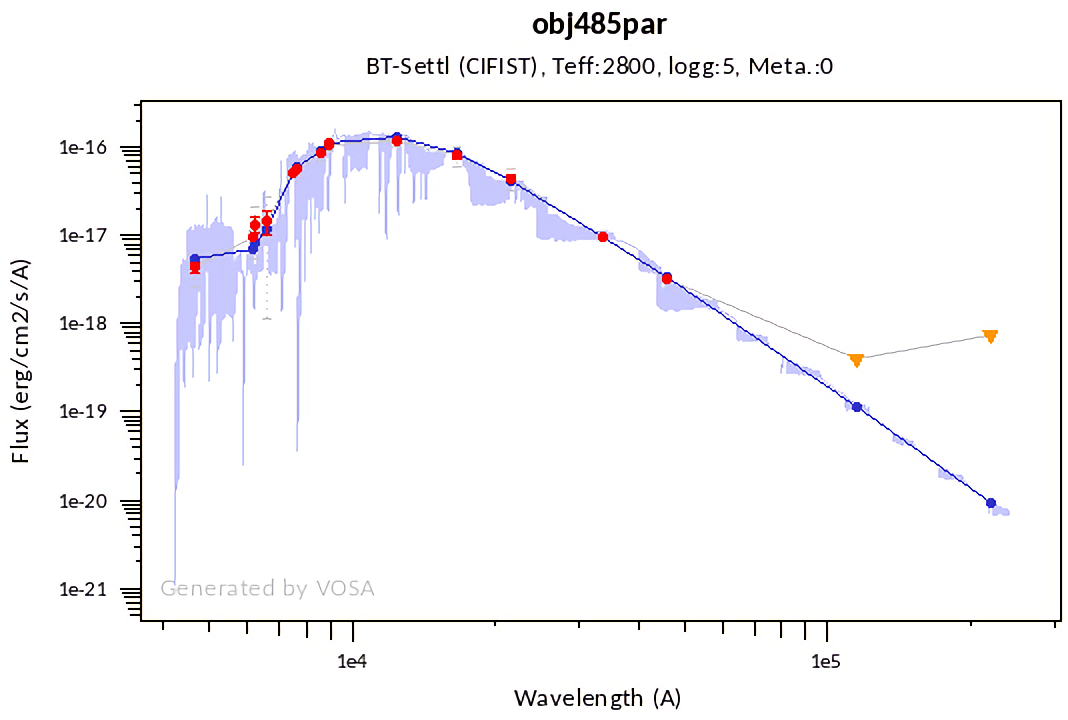}
    \caption{Example of an automatically generated SED fitting with \texttt{VOSA}. The blue spectrum represents the theoretical model that fits best, while red dots represent the observed photometry. The inverted yellow triangle indicates that the photometric value corresponds to an upper limit. These points are not considered in the fitting process.}
    \label{fig:vosa_sed}
\end{figure}

To estimate physical properties, such as effective temperature, luminosity or radius of the shortlisted objects described in the previous sections, we made use of the tool \texttt{VOSA}\footnote{\url{http://svo2.cab.inta-csic.es/theory/vosa/}} \citep{vosa}. This is a tool developed and maintained by the Spanish Virtual Observatory\footnote{\url{https://svo.cab.inta-csic.es}} which fits observational data to different collections of theoretical models. An example of \texttt{VOSA} Spectral Energy Distribution (SED) fitting can be found in Fig. \ref{fig:vosa_sed}. Before doing the fit, we built the observational SEDs using the J-PLUS photometric information as well as additional photometry from the 2MASS, UKIDSS, WISE, and VISTA infrared surveys, and from the SDSS data release 12 optical catalogue, available in \texttt{VOSA}.

In our analysis, we used the BT-Settl (CIFIST) collection of theoretical models \citep{allard2012, caffau2011}. Thus, the effective temperature estimated by \texttt{VOSA} is discretised due to the step adopted in the CIFITS grid of models (100\,K). We also assumed a surface gravity logg in the range 4.5 to 5.5 and solar metallicity. The limiting magnitude (5$\sigma$, 3\,arcsec diameter aperture) of J-PLUS DR2 is 20.5 [AB] in the \textit{z} band \citep{jpluscal}. If we take, for example, the object TVLM 891-15871, which is one of the objects in the UCD catalogue presented in \citet{Reyle2018} with the brightest absolute magnitude (11.36 [AB]) in the \textit{z} band, we see that it could be detected at a maximum distance of $\sim$680\,pc. This leads us to expect a maximum distance of about 650-700\,pc to find UCDs in the J-PLUS DR2.

Extinction plays a fundamental role in shaping the SED and, therefore, in the estimation of physical parameters \citep{extinction_laugalys, extinction_straizys}. Considering the maximum distance at which UCDs can be detected with J-PLUS, we adopted a range of values between $A_V=0$\,mag and $0.5$\,mag. We relied on the calibration described in Table 1 of \citet{solano2021} to adopt a temperature cutoff of 2\,900\,K for UCDs if the BT-Settl (CIFIST) models are used in \texttt{VOSA}. The goodness of fit of the SED in \texttt{VOSA} can be assessed with the vgfb parameter, a pseudo-reduced $\chi^2$ internally used by \texttt{VOSA} that is calculated by forcing $\sigma(F_{\rm obs}) > 0.1\times F_{\rm obs}$, where $\sigma(F_{\rm obs})$ is the error in the observed flux ($F_{\rm obs}$). Only sources with good SED fitting (vgfb < 12) were kept.

After applying these effective temperature and vgfb conditions, we used \texttt{TOPCAT} to remove the objects with a non-zero confusion flag (\texttt{cc\_flg}) in 2MASS, so as to ensure that objects are not contaminated or biased due to the proximity to a nearby source of equal or greater brightness. Moreover, we used the \texttt{Aladin} sky atlas \citep{aladin} to carry out a visual inspection of the coldest objects, in order to discard any problem related to blending or contamination by nearby objects. Finally, we ended up with 9\,810 final candidate UCDs. For the record, we checked that 204 of these objects have a renormalised unit weight error \citep[RUWE; ][]{lindegren2018} greater than 1.4 in \textit{Gaia} EDR3, which could mean that the source is affected by close binary companions. These objects were not removed since a binarity analysis is performed in Sect. \ref{binarity}.

\begin{figure}
	\includegraphics[width=\columnwidth]{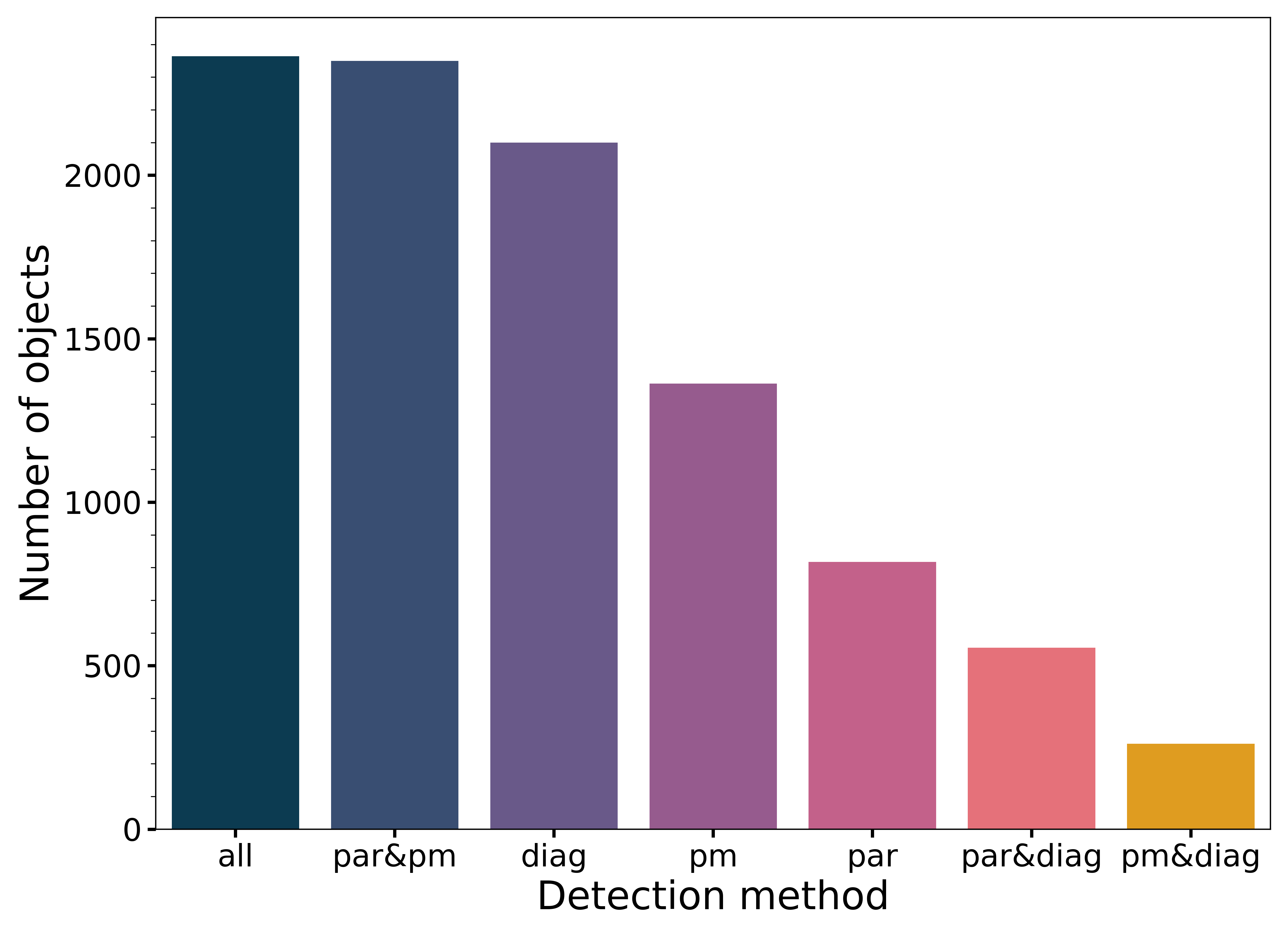}
    \caption{Breakdown of our candidate UCDs according to the methods by which they have been detected. `diag' label represents photometry-based selection, while `par' and `pm' labels represent selections based on parallax and proper motion, respectively. The `all' label comprises the candidate UCDs identified with the three approaches.}
    \label{fig:methods_both}
\end{figure}

As we use multiple detection methods in our methodology, distinct candidate UCDs may have been detected by different methods, or by several of them. Fig. \ref{fig:methods_both} shows the breakdown of the 9\,810 candidate UCDs according to the methods by which they have been detected. The fact that 2\,100 objects are only detected by the photometric methodology (`diag' bar in Fig. \ref{fig:methods_both}) and 4\,530 are only detected by the astrometric methodology (`par', `pm', and `par\&pm' bars in Fig. \ref{fig:methods_both}) argues for the complementary nature of both approaches. Considering each method separately, we detected 6\,086 candidates with parallax-based selection, 6\,338 with proper motion-based selection, and 5\,280 with photometry-based selection.

%--------------------------------------------------------------------

\section{Analysis} \label{Analysis}

\subsection{Temperatures and distances} \label{dist_temps}

\begin{figure}
	\includegraphics[width=\columnwidth]{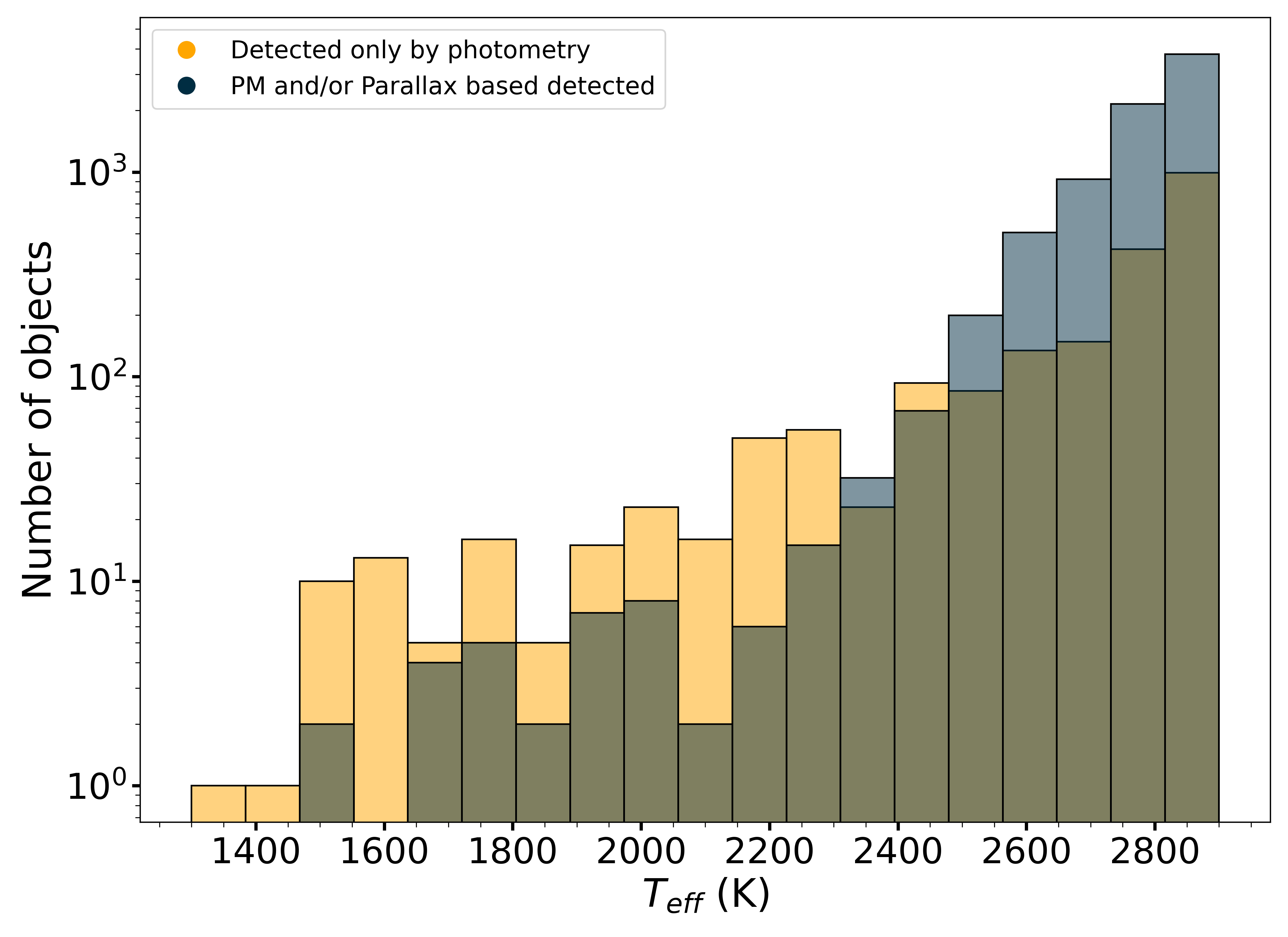}
    \caption{$T_{\rm eff}$ distribution for our candidate UCDs. In yellow we show the candidates that were only detected by photometry. In blue we show the candidates that were, at least, detected by astrometric methodology.}
    \label{fig:temphist}
\end{figure}

Fig. \ref{fig:temphist} shows that the distribution of effective temperatures for our candidate UCDs is not the same depending on whether they have been detected by astrometric methodology (blue) or not. To prove this, we performed a two-sample Kolgomorov-Smirnov test on the two samples, which returned a $p$ value = 3.66\,$\cdot\,10^{-15}$, rejecting the possibility that both samples are coming from the exact same distribution. The number of cold objects ($T_{\rm eff}\leq$ 2\,200\,K) is clearly higher in the only-photometry detected distribution (yellow). Most of our candidates ($\sim$86\,\%) have $T_{\rm eff}\geq$ 2\,700\,K, a clear consequence of the working wavelength, since UCDs peak in the near-infared, and J-PLUS covers only up to the z filter ($\lambda_{\rm eff}=8\,940.28$\,\AA).

\begin{figure}
	\includegraphics[width=\columnwidth]{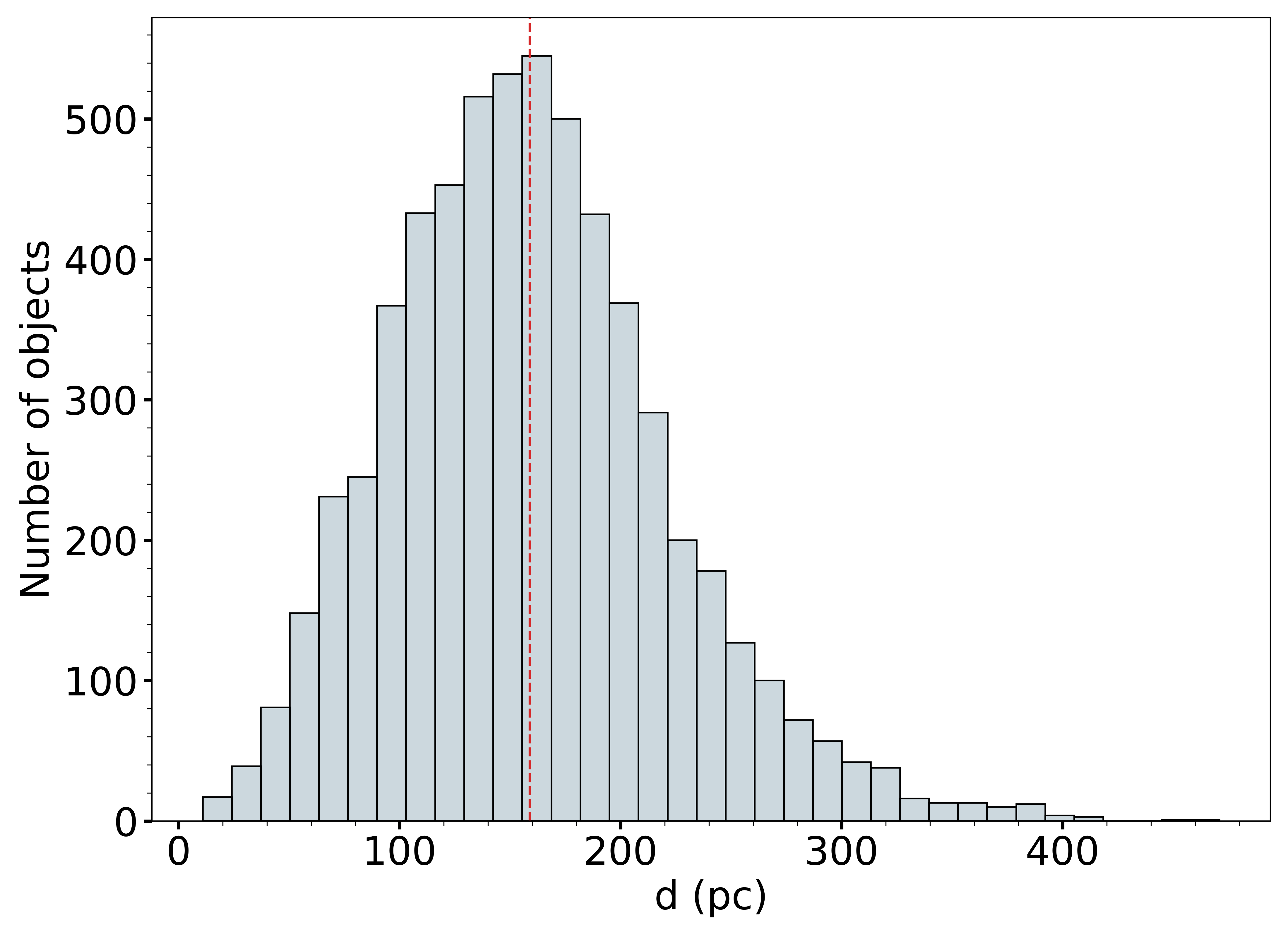}
    \caption{Distance distribution for our candidate UCDs with an error of less than 20\,\% in \textit{Gaia} EDR3 J2016 parallax. The mean value (red vertical line) of the distribution is 159\,pc, with the closest and farthest objects at 11\,pc and 471\,pc.}
    \label{fig:disthist}
\end{figure}

\begin{figure}
	\includegraphics[width=\columnwidth]{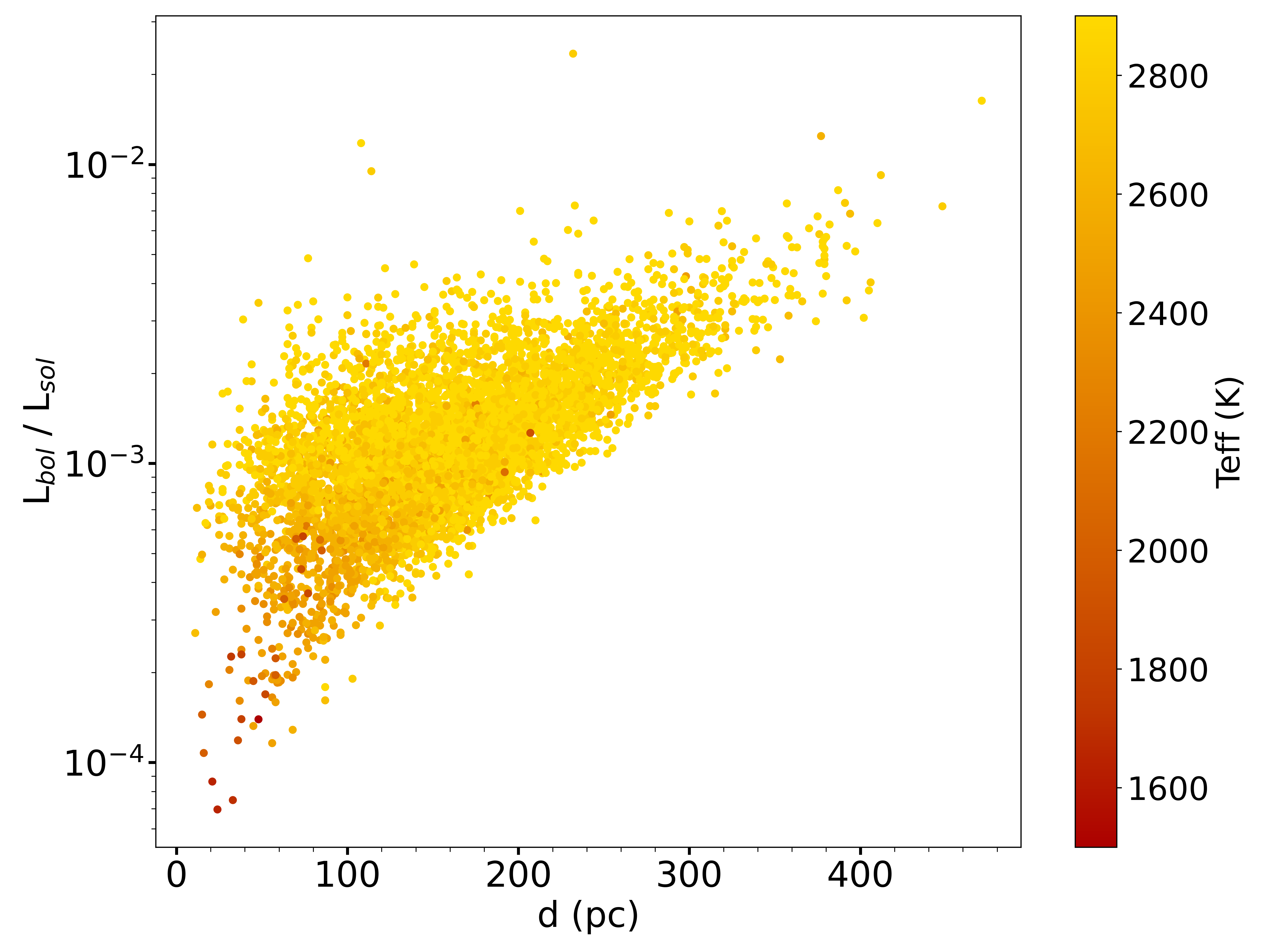}
    \caption{Bolometric luminosity (in solar units) vs. distance diagram of our candidate UCDs with good parallaxes. The points are colour-coded by temperature.}
    \label{fig:temp_lum}
\end{figure}

For the distance distribution of our candidate UCDs (Fig. \ref{fig:disthist}), we only considered the candidates with a relative error of less than 20\,\% in parallax (6\,086 objects), so we can rely on the inverse of the parallax as a distance estimator \citep{Luri2018}. In our case, as mentioned in Sect. \ref{Methodology}, the parallax are those of \textit{Gaia} EDR3. About 70\,\% of the objects lie in the $96 < {\rm D\,(pc)} < 222$ region (1$\sigma$ limits), with a maximum and minimun distance of 471\,pc and 11\,pc, respectively. This upper limit is consistent with the value estimated in Sect. \ref{vosa}. We found 68 nearby objects, at distances smaller than 40\,pc, that will be further discused in Sect. \ref{new_vs_known}. Fig. \ref{fig:temp_lum} gives a more in-depth view of the characteristics of our candidate UCDs. As expected, most of the cooler candidates are detected at closer distances and tend to have lower bolometric luminosity.

\subsection{Kinematics} \label{tangential_vel}

\begin{figure}
	\includegraphics[width=\columnwidth]{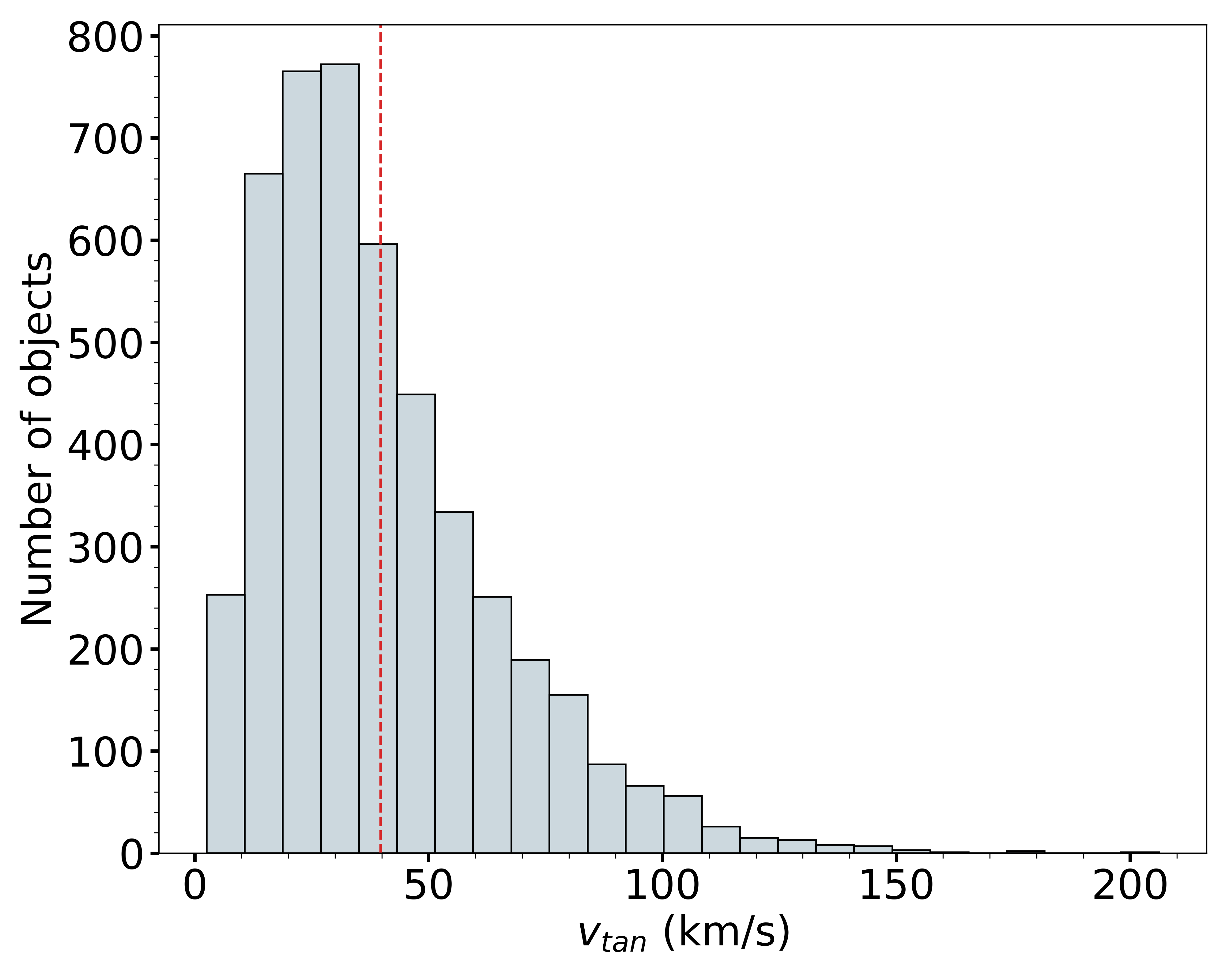}
    \caption{Tangential velocity distribution for our candidate UCDs with good parallax and pm conditions. The mean value (red vertical line) of the distribution is 39.78\,km\,s$^{-1}$, with a maximum value of 206\,km\,s$^{-1}$.}
    \label{fig:vtan_hist}
\end{figure}

Stellar kinematics is a reliable proxy for segregating large-scale galactic populations (thin disk, thick disk, and halo) \citep{Burgasser2015}. Using \textit{Gaia} EDR3 proper motions and parallaxes, we computed the tangential velocities of our candidate UCDs as $v_{\rm tan}=4.74 \mu d$, where $v_{\rm tan}$ is given in km\,s$^{-1}$, $\mu$ is the total proper motion in arcsec\,yr$^{-1}$ and $d$ is the distance in pc. For a correct estimation of the tangential velocity, we only considered candidates that met both conditions described in Sects. \ref{parallax} and \ref{pm} for good parallax and proper motion (4\,714). Fig. \ref{fig:vtan_hist} shows the distribution of tangential velocities for these candidates, with  a mean value of $v_{\rm tan}=39.78$\,km\,s$^{-1}$, a median value of $v_{\rm tan}=33.99$\,km\,s$^{-1}$, and a dispersion of $\sigma_{\rm tan}=24.85$\,km\,s$^{-1}$. Even taking into account objects located at the long tail of the distribution (134 objects, representing 2.8\,\% of the total, with $v_{\rm tan}>100$\,km\,s$^{-1}$), these values agree with previous calculations for UCDs \citep{faherty2009}.

\citet[][Fig. 10]{Torres2019} shows a breakdown of the tangential velocity based on the membership in the thin disk, the thick disk or the halo. Relying on these values, we can segregate our candidate UCDs into thin disk ($v_{\rm tan}\leq 85$\,km\,s$^{-1}$), thick disk ($85< v_{\rm tan}< 155$\,km\,s$^{-1}$), and halo ($v_{\rm tan}\geq 155$\,km\,s$^{-1}$) populations. We found 4\,441, 268 and five candidate UCDs in these intervals, respectively. According to \citet{Kilic2017}, the corresponding ages are 6.8-7.0\,Gyr (thin disk), 7.4-8.2\,Gyr (thick disk), and 12.5$^{+1.4}_{-3.4}$\,Gyr (halo).

\begin{figure}
	\includegraphics[width=\columnwidth]{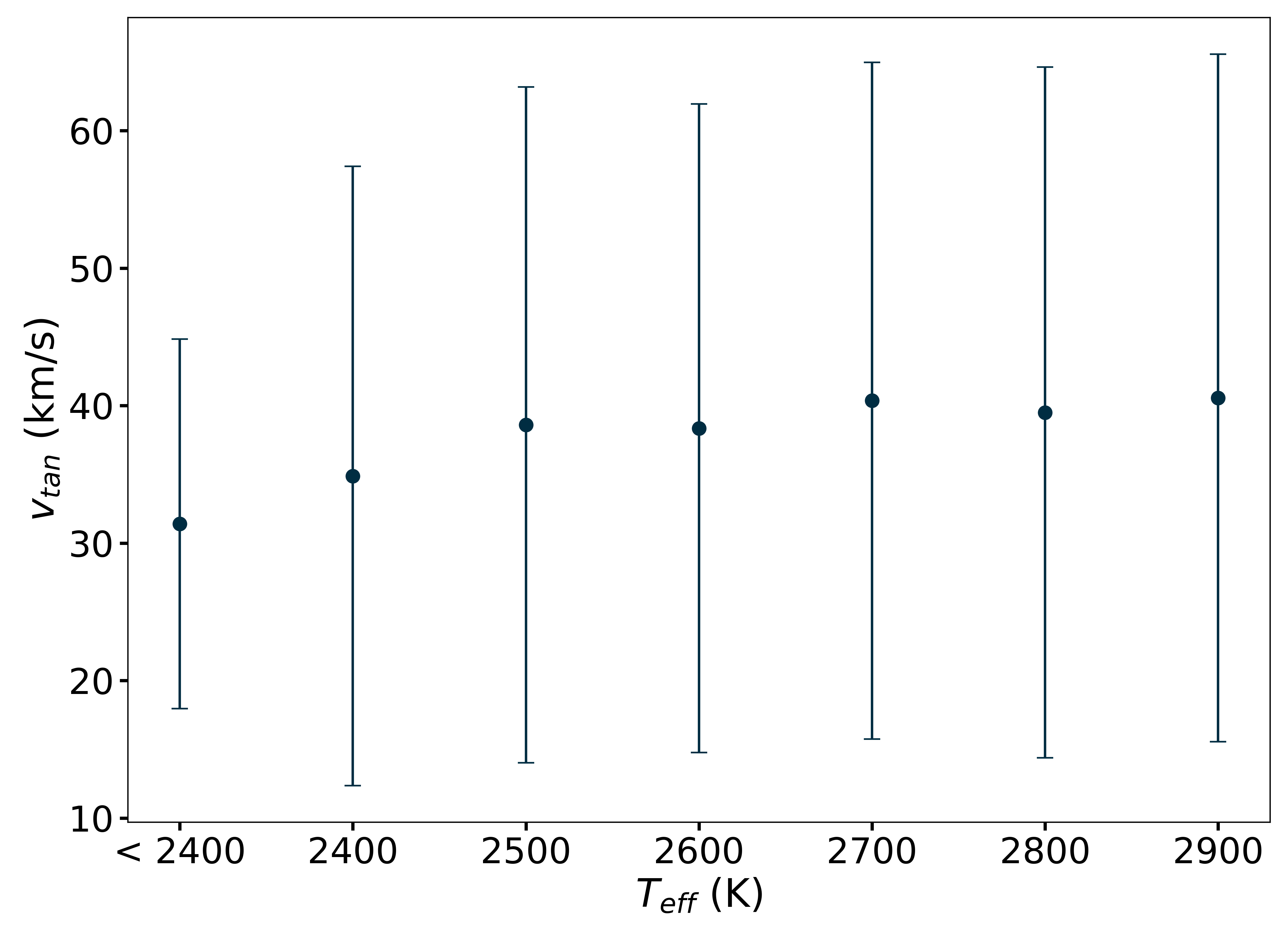}
    \caption{Mean tangential velocity for each value of effective temperature of our candidate UCDs with reliable parallax and proper motion. The error bars represent the standard deviation.}
    \label{fig:vtan_temp}
\end{figure}

Three of the potential halo members show a very high tangential velocity. Two of them, with Simbad identifiers 2MASS J18030236+7557587 and 2MASS J13155851+2814524, are not far from the thick disk-halo threshold, with tangential velocities of $v_{\rm tan}=176.25$\,km\,s$^{-1}$ and $v_{\rm tan}=177.47$\,km\,s$^{-1}$, respectively. Furthermore, one of the objects has $v_{\rm tan}=206.16$\,km\,s$^{-1}$, which significantly exceeds the limit. This object, at a distance of 179 pc, is reported as an M7 in the catalogue provided by \citet{ahmed2019} with the id J132625.03+333506.7. Due to its high tangential velocity, we conclude this object could be a potential member of the Galactic halo. We used the ($J-K_{\rm s}$, $i-J$) colour-colour diagram presented in \citet{lodieu2017} to study the metallicity of this object. With values of $J-K_{\rm s}=0.77$ and $i-J=3.29$, the object exhibits subdwarf behaviour (low metallicity). Fig. \ref{fig:vtan_temp} shows the mean and standard deviation of the tangential velocity for each value of the effective temperature. There is no evidence of correlation between effective temperature and tangential velocity among our candidates. 

To study the possible membership of our candidate UCDs to nearby young associatons, we relied on \texttt{BANYAN}~$\Sigma$\footnote{\url{http://www.exoplanetes.umontreal.ca/banyan/}} \citep{banyan}, a Bayesian analysis tool to identify members of young associations. Modelled with multivariate Gaussians in six-dimensional $\rm XYZUVW$ space, \texttt{BANYAN}~$\Sigma$ can derive membership probabilities for all known and well-characterised young associations within 150\,pc. As we found no radial velocity data available for any of the 4\,714 candidate UCDs with good parallax and proper motion, we introduced the sky coordinates, proper motion, and parallax of these objects as input parameters to the algorithm. 

For 4\,666 of the candidate UCDs, the algorithm predicted that most of them are field stars. However, it gave a high Bayesian probability for 48 objects to belong to a young association, in 30 of the cases with a probability greater than 95\,\%. In more detail, the algorithm mapped 34 candidate UCDs to the Pisces-Eridanus stellar stream \citep{PERI}, five to the Argus Association \citep{argus}, four to the AB Doradus Moving Group \citep{abdmg}, two to the Columba association \citep{columba}, and one each to the Tucana-Horologium \citep{tha}, $\beta$ Pictoris \citep{bpictoris}, and Carina-Near \citep{carn} associations. We verified all these 48 objects have tangential velocities typical of the thin disk, with mean $v_{\rm tan}=16.37$\,km\,s$^{-1}$ and standard deviation $\sigma = 6.17$\,km\,s$^{-1}$. As mentioned in \texttt{BANYAN}~$\Sigma$, a high membership probability in a young association does not guarantee that the star is a true member, or young, so further follow-up would be needed to demonstrate the youth of the object. 

\subsection{Binarity} \label{binarity}

We conducted a search for binary systems among our candidate UCDs in two ways. We searched for unresolved binaries using a methodology based purely on the photometry of our objects. Using the complementary photometry functionality of \texttt{VOSA}, we selected only the candidates fulfilling three conditions. First, with an excess detected by \texttt{VOSA} in any filter in the infrared. We discarded WISE $W3$ and $W4$ due to their poor angular resolution and sensitivity. Second, with good photometry in both 2MASS (\texttt{Qfl} = A) and WISE (\texttt{cc\_flags} = 0 and \texttt{ph\_qual} = A or B). Third, with at least three good photometric points in the infrared, apart from the detected excess.

After applying these conditions, we ended up with 291 objects with an excess in the infrared that could be ascribed to circumstellar material or to the presence of a close ultracool companion. Then, we used the binary fit functionality of \texttt{VOSA} to fit the observed SED of these 291 objects using the linear combination of two theoretical models. After this, we ended up with 122 candidate UCDs for which the infrared excess detected is nicely reproduced by performing a two-body fit, suggesting the existence of an unresolved companion.

In parallel to this, we looked for \textit{Gaia} companions of our candidate UCDs at large angular separations, using only those with reliable parallax and proper motion (4\,714). Firstly, we cross-matched these sources with \textit{Gaia} EDR3 J2016 to get all the objects separated a maximum of 180 arcsec in the sky (maximum separation allowed by the \texttt{X-match} service in \texttt{TOPCAT}) from each of our candidate UCDs. Then, we established a conservative upper limit of 100\,000\,au for the projected physical separation between a candidate and its companion. Finally, we relied on the conditions presented in \citet{smart_dwarfs} to ensure that the companion shares a parallax and proper motion similar to that of our candidate UCD:

\begin{itemize}

    \item $\Delta \varpi < max[1.0, 3\sigma_{\rm \varpi}]$  
    \item $\Delta(\mu_{\alpha}\cos{\delta}) < 0.1\mu_{\alpha}\cos{\delta}$ 
    \item $\Delta \mu_{\delta} < 0.1\mu_{\delta}$ 

\end{itemize}

\noindent where $\varpi$ and $\mu$ are the parallax and proper motion of our candidate UCDs, respectively. After applying these criteria, we ended up with 73 candidate UCDs with one \textit{Gaia} companion and another five candidate UCDs with two \textit{Gaia} companions identified. Of these 78 objects, six are already tabulated as known binary systems by the Washington Double Star catalogue \citep[WDS;][]{WDS}. Table \ref{tab:binaries} lists the coordinates (J2000), parallaxes, proper motions, angular separations $\rho$ and projected physical separations $s$ of the six known systems. A table with the same information for the identified multiple systems that are not tabulated by the WDS is accesible through the catalogue described in Appendix \ref{vo_catalogue}.

\begin{table*}
 \caption{Identified binary systems that are already tabulated by the WDS.}
 \label{tab:binaries}
 \centering          
 \begin{tabular}{l c c c c c c c}
  \hline\hline
  \noalign{\smallskip}
  
  Name\,$^{(a)}$ & Spectral & $\alpha$ & $\delta$ & $\varpi$ & $\mu_{\rm tot}$ & $\rho$ & $s$\\
    & type\,$^{(b)}$ & (deg) & (deg) & (mas) & (mas\,yr$^{-1}$) & (arcsec) & (au)\\
  
  \noalign{\smallskip}
  \hline
  \noalign{\smallskip}

  2MASS J01560053+0528562 & M6 & 29.00219 & 5.48230 & $11.66\pm0.09$ & $153.33\pm0.14$ & 7.3 & 621.4 \\
  2MASS J01560037+0528494 & M7.5 & 29.00194 & 5.47979 & $11.74\pm0.21$ & $154.39\pm0.32$ &  &  \\  

  \noalign{\smallskip}
  
  LSPM J0209+0732 & \ldots & 32.39985 & 7.54052 & $8.27\pm0.08$ & $166.00\pm0.11$ & 27.1 & 3881.2 \\
  2MASS J02093416+0732196 & \ldots & 32.39316 & 7.53862 & $6.99\pm1.00$ & $166.29\pm1.82$ &  &  \\ 

  \noalign{\smallskip}
  
  SLW J0851+4134A & M4.5 & 132.94640 & 41.57082 & $21.84\pm0.04$ & $114.65\pm0.05$ & 32.7 & 1503.1 \\
  2MASS J08514823+4134453 & M7\,V & 132.95016 & 41.57936 & $21.69\pm0.18$ & $114.10\pm0.24$ &  &  \\
  
  \noalign{\smallskip}
  
  Gaia EDR3 803883569892552320\,$^{(c)}$ & \ldots & 151.67670 & 41.58267 & $5.94\pm0.13$ & $52.53\pm0.15$ & 18.3 & 3110.5 \\
  SDSS J100640.86+413503.9 & M6\,V & 151.67006 & 41.58431 & $5.88\pm0.21$ & $52.66\pm0.23$ &  &  \\ 
  
  \noalign{\smallskip}
  
  SDSS J164331.99+634340.6 & \ldots & 250.88343 & 63.72784 & $13.55\pm0.12$ & $74.66\pm0.20$ & 5.8 & 415.1 \\
  Gaia EDR3 1631815065395291392\,$^{(c)}$ & \ldots & 250.88663 & 63.72773 & $13.71\pm0.12$ & $75.58\pm0.20$ &  &  \\

  \noalign{\smallskip}

  SLW J1840+4204A & M7.9 & 280.23178 & 42.06685 & $14.11\pm0.11$ & $59.67\pm0.19$ & 17.4 & 1217.5 \\
  SLW J1840+4204B & M8.1 & 280.23762 & 42.06474 & $14.59\pm0.39$ & $59.75\pm0.68$ &  &  \\ 
  
  \noalign{\smallskip}
  \hline
 \end{tabular}
  \tablefoot{$^{(a)}$ Primaries and secondaries are sorted as in the WDS. $^{(b)}$ When it is not available in Simbad, we show the spectral type given in the WDS. $^{(c)}$ The \textit{Gaia} source id is displayed, since the object is not reported in Simbad.}
\end{table*}

A deeper knowledge of the \textit{Gaia} companion may allow us to infer properties, such as metallicity, of our candidate UCD. We only found spectral types in Simbad for two of the detected companions, with spectral types F2 and K3V. To obtain information about the rest of the companions, we first made use of \texttt{VOSA} to get an estimate of their effective temperature. Then, we relied on the updated version of Table 5 in \citet{pecaut2013} to map these effectives temperatures to the spectral types of the companions. As result, we ended up with four F-type, one G-type, 16 K-type and 42 M-type stars among the companions with good SED fitting in \texttt{VOSA}. For the rest of the companions, we obtained a bad SED fitting in \texttt{VOSA} (vgfb > 12), so we could not get an estimation of the effective temperature.

%--------------------------------------------------------------------

\section{Known ultracool dwarfs} \label{known}

\subsection{Recovered known UCDs} \label{recovered_known}
 
Here, we assess the number of known UCDs found in the J-PLUS DR2 field and the fraction of them that were recovered using our methodology. For this analysis, we used nine catalogues and services: SIMBAD\footnote{\url{http://simbad.u-strasbg.fr/simbad/}}\citep{Wenger00}, \citet{zhang2009}, \citet{zhang2010}, \citet{schmidt2010}. \citet{skrzypek2016}, \citet{smart2017}, \citet{Reyle2018}, \citet{panstarrs1}, and \citet{ahmed2019}. Using the SIMBAD TAP service\footnote{\url{http://simbad.u-strasbg.fr:80/simbad/sim-tap}} through \texttt{TOPCAT}, we selected objects with spectral types M7\,V, M8\,V, M9\,V or labelled as brown dwarfs. A total of 18\,282 objects were recovered. Also, from \citet{panstarrs1} we chose the 2\,090 objects having spectral type M7 or later. As all the 33\,665, 14\,915, 1\,886, 1\,361, 806, 484 and 129 objects in the \citet{ahmed2019}, \citet{Reyle2018}, \citet{smart2017}, \citet{skrzypek2016}, \citet{zhang2010}, \citet{schmidt2010}, and \citet{zhang2009} catalogues, respectively, are within our scope (spectral type M7 or later), we included them in their entirety. 

To select only the known UCDs that lie in the region of the sky covered by J-PLUS DR2 we made use of \texttt{TOPCAT} and its \texttt{nearMOC} functionality, which indicates whether a given sky position either falls within, or is within a certain distance of the edge of, a given MOC. The MOC\footnote{\url{https://www.ivoa.net/documents/MOC/}} (Multi-Order Coverage Map) is an encoding method dedicated to VO applications or data servers which allows to manage and manipulate any region of the sky, defining it by a subset of regular sky tessellation using the HEALPix method \citep{healpix}. Out of a total of 5\,817 objects lying in the J-PLUS DR2 field of view, we ended up with 4\,734 known UCDs with photometry in the relevant J-PLUS filters described in Sect. \ref{Methodology} (see Table \ref{tab:filters}), which are reduced to 4\,649 objects after removing those with non-zero confusion and contamination flags in 2MASS. From this set, 1\,983 were recovered using our methodology and 2\,666 were not. We conducted an in-depth analysis of the 2\,666 UCDs following the two methodologies (astrometric and photometric) separately, to see in which steps of the process these objects are discarded.

In short, of this 2\,666 unrecovered objects, 1\,520 are lost because they do not meet our parallax, proper motion, and photometry constraints, while another 119 are discarded in the $G - G_{\rm RP}$ and $r - z$ cuts. The remaining 1\,027 are lost in the temperature/vgfb cutoff after the analysis with \texttt{VOSA}, some due to a bad SED fitting (vgfb > 12) and most of them due to an estimated temperature higher than 2\,900\,K. We have checked the latter and the vast mayority of them are M7\,V from Simbad that lie at the temperature limit, with estimated temperatures of 3\,000 - 3\,100\,K,

\subsection{New candidate UCDs vs. previously known} \label{new_vs_known}

In this section, we analyse the differences between previously known UCDs and the remaining candidate UCDs among our sample. For this, we cross-matched our candidate UCDs with the known UCDs sample described in Sect. \ref{known}. As indicated above, of the 9\,810 candidates identified by the proposed VO methodology, only 1\,983 were previously reported as UCD. This amounts to a total of 7\,827 new candidate UCDs in the sky coverage of J-PLUS DR2, which represents an increase of about 135\,\% (7\,827/5\,817) in the number of UCDs for this area.

\begin{figure}
	\includegraphics[width=\columnwidth]{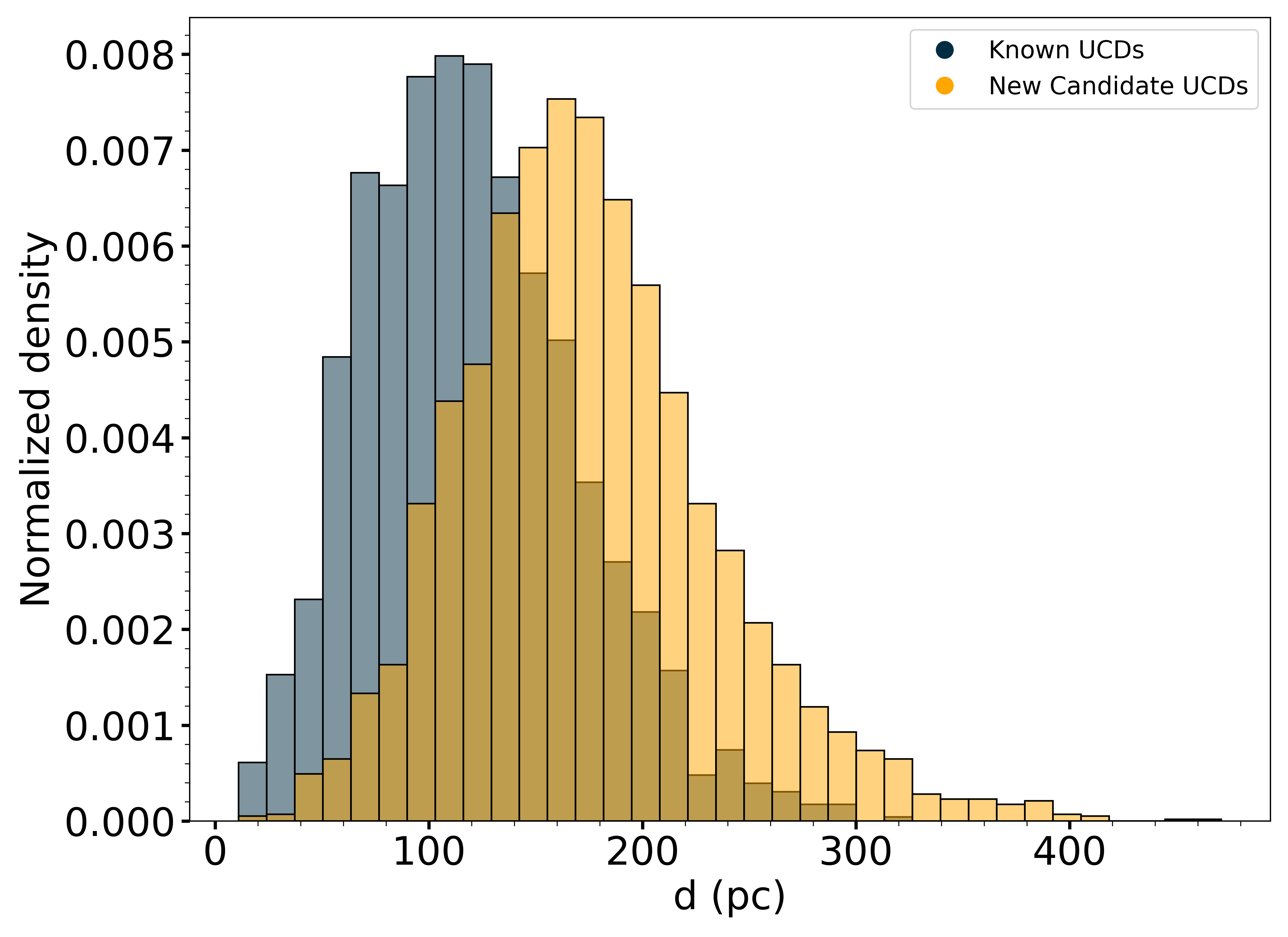}
    \caption{Distance distribution for previously reported (blue) and new (yellow) candidate UCDs with good parallax conditions.}
    \label{fig:disthistboth}
\end{figure}

\begin{figure}
	\includegraphics[width=\columnwidth]{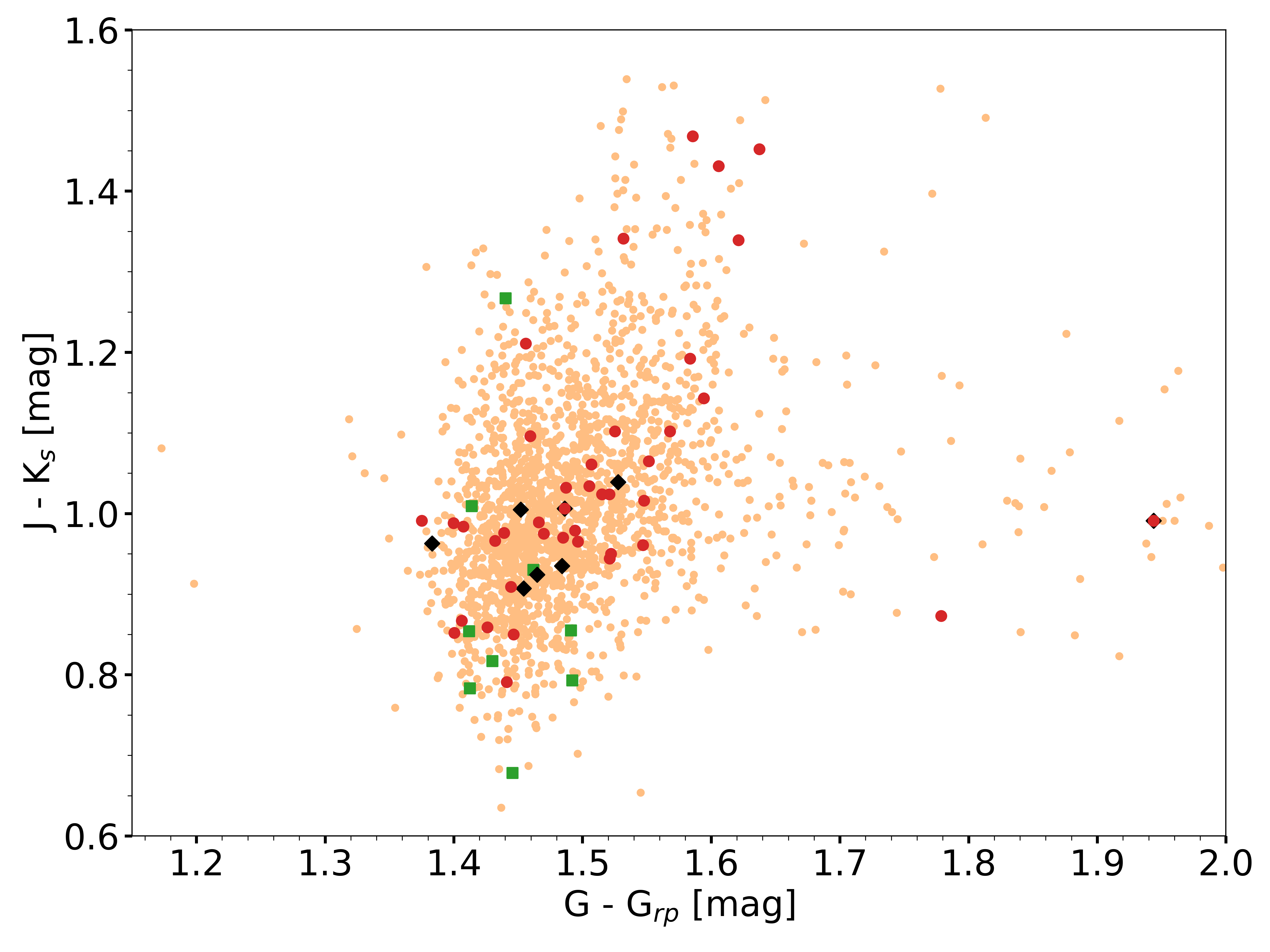}
    \caption{\textit{$J-K_s$} (2MASS) vs. \textit{$G-G_{\rm RP}$} (\textit{Gaia}) diagram of our candidate UCDs with good 2MASS photometric quality (Qflg=A) in \textit{J} and \textit{$K_s$} bands. Black diamonds represent our eight new nearby candidate UCDs at distances $d<40$\,pc. Green squares stand for new candidate UCDs with tangential velocities $v_{\rm tan}>100$
    \,kms\,$^{-1}$. Red circles represent candidate UCDs with a possible membership in a nearby young association.}
    \label{fig:2massdiag}
\end{figure}

Fig. \ref{fig:disthistboth} shows the distance distribution for our candidate UCDs, with good parallax conditions, discriminated by colour according to whether or not they were previously reported as UCD. It is clear that the new candidates detected are, on average, more distant, driven by the improvement of the quality of parallaxes with \textit{Gaia} EDR3. Of the 68 nearby objects found at distances smaller than 40\,pc, eight have not been previously reported as UCD. To check whether these objects could have been missed by other photometric surveys due to anomalies in their colours, we constructed a colour-colour diagram using \textit{$J-K_{\rm s}$} (2MASS) and \textit{$G-G_{\rm RP}$} (\textit{Gaia}) colours. Fig. \ref{fig:2massdiag} shows that this is not the case for any of these objects (black diamonds in the diagram).

\begin{figure}
	\includegraphics[width=\columnwidth]{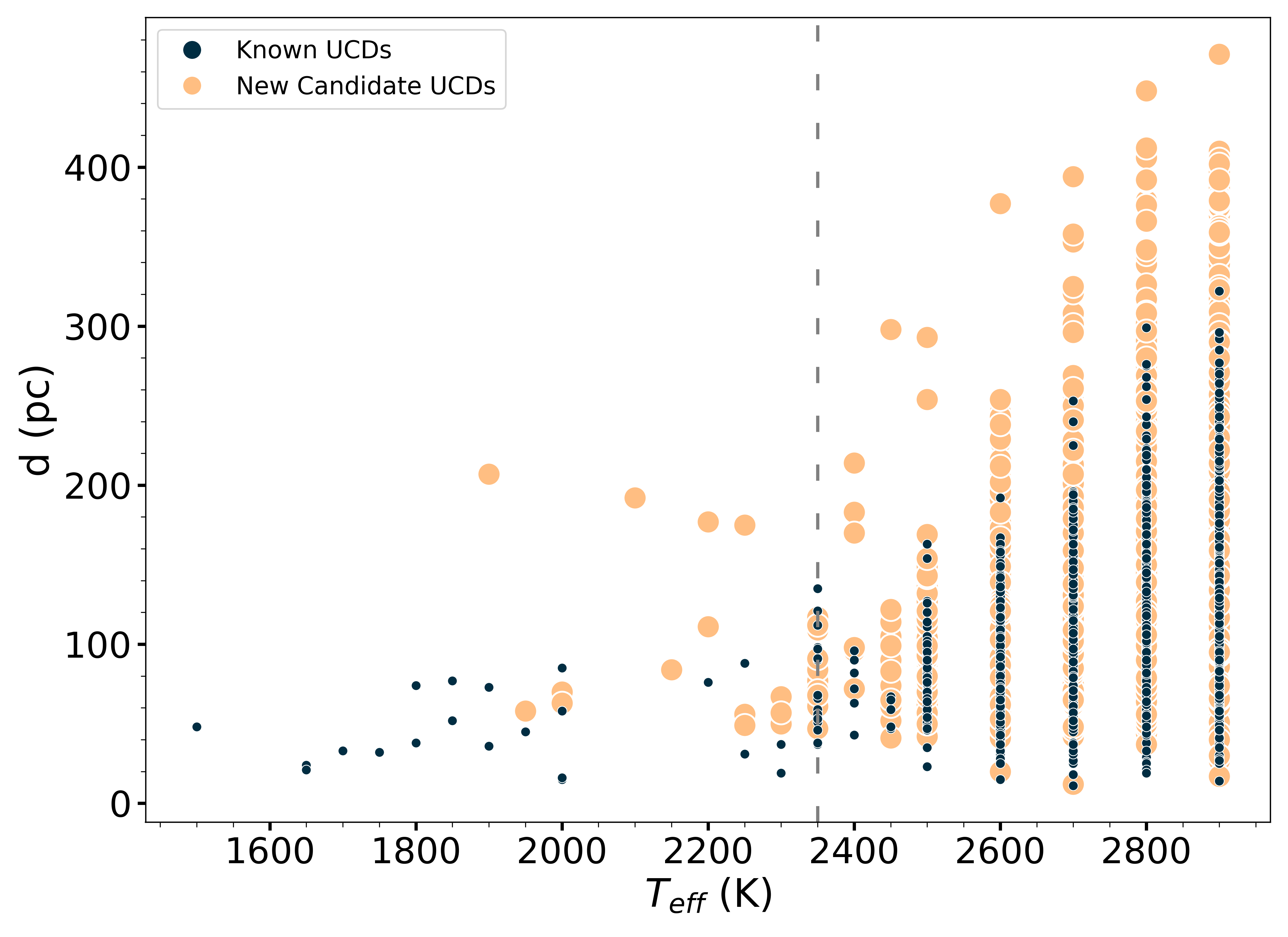}
    \caption{Distance vs. effective temperature diagram for previously reported (blue) and new (yellow) candidate UCDs with good parallax conditions. The vertical dashed line indicates the lower limit of effective temperature for M-type dwarfs (2\,359\,K) according to \citet{pecaut2013}.}
    \label{fig:distteff}
\end{figure}

A more in-depth view of this is the distance vs. effective temperature diagram shown in Fig. \ref{fig:distteff}. Here we can see how previously reported candidate UCDs tend to be at shorter distances for any value of the effective temperature. This trend is more clearly observed for higher temperature values, where the diagram shows how the new candidate UCDs cover the range of distances of the previously reported candidates and extend it to larger values, suggesting that our methodology allows us to go further in the search for new UCDs.

\begin{figure}
	\includegraphics[width=\columnwidth]{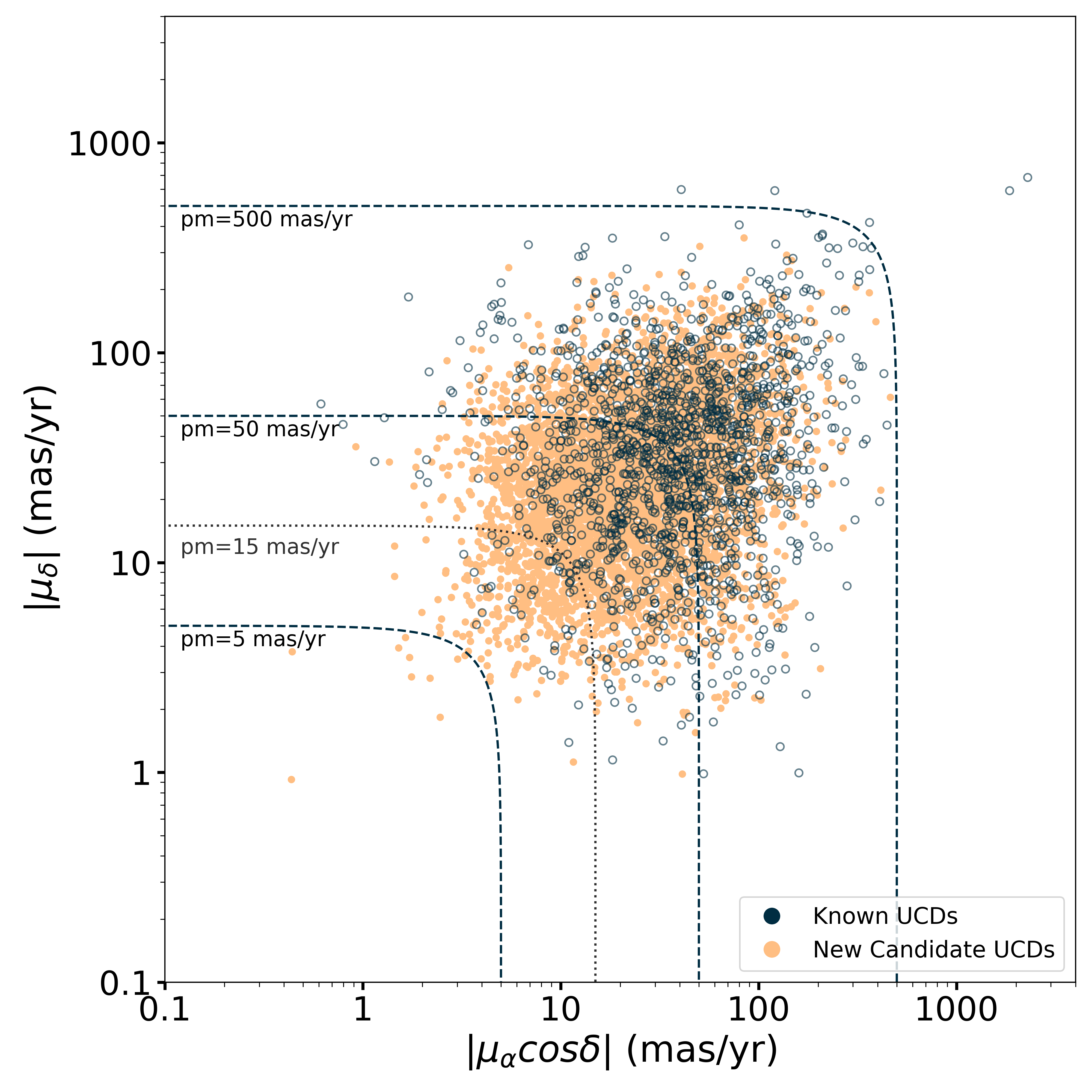}
    \caption{Absolute proper motion components for previously reported (blue) and new (yellow) candidate UCDs with good proper motion conditions.}
    \label{fig:pmanalysis}
\end{figure}

Going further, in Fig. \ref{fig:pmanalysis} we plot the absolute proper motions |$\mu_{\delta}$| and |$\mu_{\alpha}\cos{\delta}$| for our candidate UCDs with good proper motion conditions. It shows how the new candidate UCDs detected extend to smaller values of proper motion. Especially for values of proper motion of less than $15$\,mas\,yr$^{-1}$, the number of new candidates is significantly higher than the number of previously reported candidate UCDs, which reflects the improvement of the quality of proper motions with \textit{Gaia} EDR3.

%--------------------------------------------------------------------

\section{Machine learning analysis} \label{ml}

The filter system of J-PLUS offers a sufficiently high-dimensional space to reliably use ML techniques. We explored the ability to reproduce the presented search for candidate UCDs with a purely ML-based methodology that uses only J-PLUS photometry. Because the sample is strongly imbalanced, as a first step in the candidate UCDs identification, we proposed a filtering strategy to discard the objects that differ the most from the UCDs using the PCA algorithm. Then, with the reduced sample, SVM models were trained and fine-tuned to maximise the identification of candidate UCDs.

Principal component analysis~\citep{hotelling:33}, one of the most popular linear dimensionality reduction algorithms, is a non-parametric method that aims to reduce a complex data set to a lower dimension by identifying the axes that account for the largest amount of variance. The unit vectors defining each of these axes are called principal components. PCA works on the assumption that principal components with larger associated variance encompass the underlying structure of the data set in order to find the best basis for re-expressing it. The expectation behind this method, as with any method of dimensionality reduction, is that the entire data set can be well characterised along a small number of dimensions (principal components). By projecting the data set onto the hyperplane defined by these principal components, you ensure that the projection will preserve as much variance as possible. 

The selection of PCA in our approach instead of other non-linear dimensionality reduction techniques, such as t-Distributed Stochastic Neighbor Embedding~\citep[t-SNE;][]{vandermaaten08} or Uniform Manifold Approximation and Projection~\citep[UMAP;][]{McInnes2018}, is mainly based on (1) the computational efficiency, since PCA allows projecting new data along the new axes without having to reapply the algorithm, and (2) the deterministic nature of the PCA solution, i.e., different runs of PCA on a given dataset will always produce the same results. These properties of PCA are crucial in our proposal, since we use the 2D representation of PCA to perform the filtering as a first step in our ML task.

Support vector machine is a supervised (requires labelled training data) ML algorithm that has been widely used in classification and regression problems~\citep{2013A&A...550A.120S,2017MNRAS.465.4556G}. The origin of this algorithm dates back to the late 70s, when \citet{vapnik} delved into the statistical learning theory. The idea behind SVM is to find a hyperplane that separates data into two classes while maximising a marging, defined as the distance from the hyperplane to the closest point across both classes. Thus, the SVM chooses the best separating hyperplane as the one that maximises the distance to these points, so the decision surface is fully specified by a subset of points on the inner edge of each class, known as support vectors. The SVM is a linear classifier, so if the data is not linearly separable in the instance space, we can gain linear separation by mapping the data to a higher dimensional space. To do so, different kernels are used, such as the polynomial or the radial basis function (RBF), since the kernel trick allows us to define a high-dimensional feature space without actually storing these features.

\subsection{PCA cut} \label{pca_cut}

In our methodology, we used J-PLUS DR2 data from one of the 20$\times$20\,deg$^2$ mentioned in Sect. \ref{Methodology}. We selected as features seven different J-PLUS colours built with the most relevant filters for UCDs, i.e., the reddest ones (see Table \ref{tab:filters}): $i-z$, $r-i$, $i-J0861$, $J0861-z$, $(i-z)^2$, $(r-i)^2$, and $r-z$. We discarded the filter \textit{J0660} because the available photometry in this filter is less abundant than in the others. Thus, we first built these variables from the J-PLUS photometry, discarding objects with no information in any of the required filters, and labelled the instances as positive or negative class using the candidate UCDs obtained with the previous methodology. After this, we ended up with a sample composed of 317 UCDs and 495\,274 non-UCD objects.

\begin{figure}
	\includegraphics[width=\columnwidth]{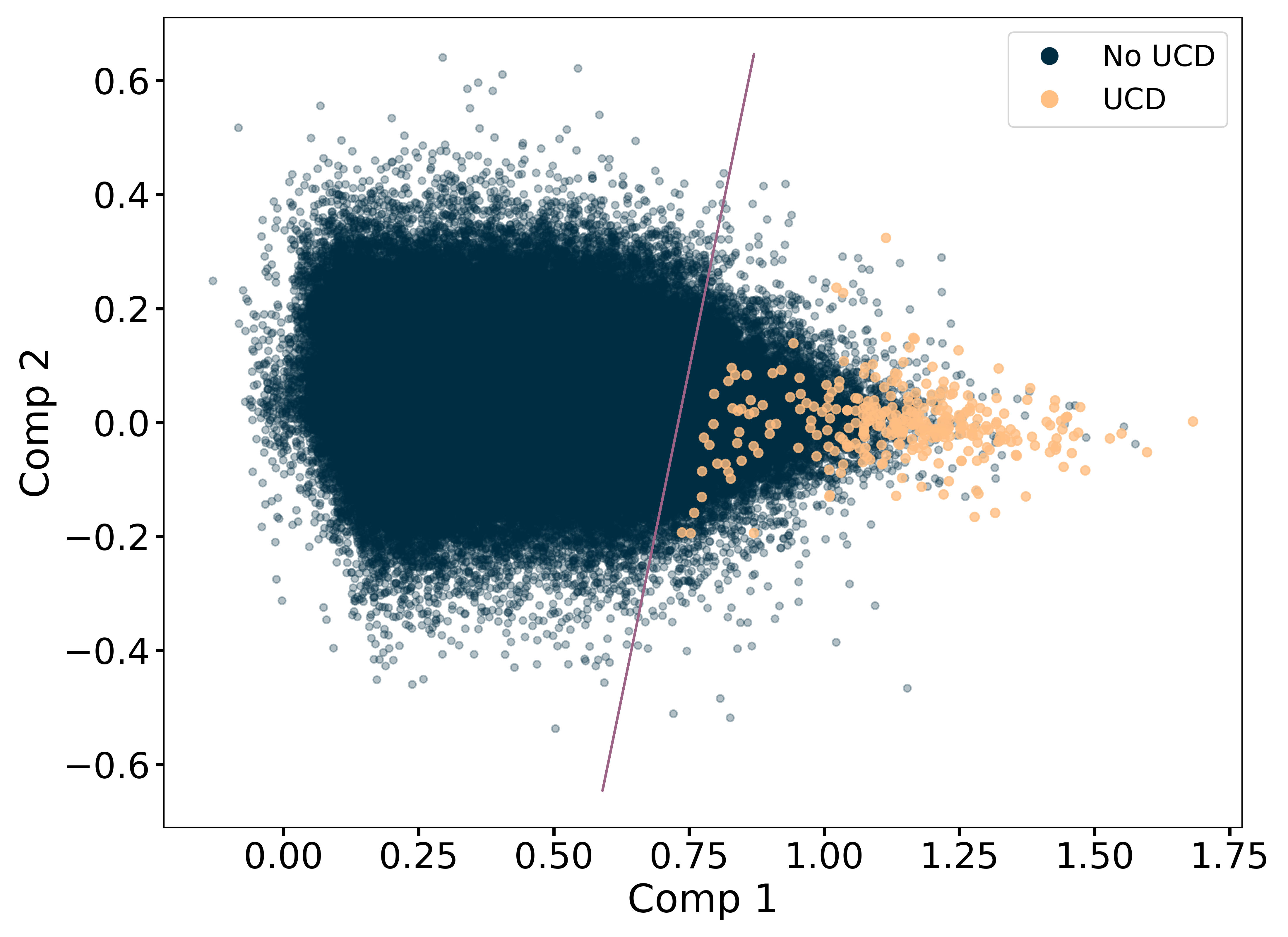}
    \caption{Projection of the sample used in the ML methodology onto the hyperplane defined by the first two principal components, with an explained variance ratio of 93\,\%. Points are colour-coded according to their class, UCD (yellow) or non-UCD (dark blue). The purple line represents the decision threshold used to make a first cut at identifying UCDs, keeping only the objects that fall on the UCD side.}
    \label{fig:pca}
\end{figure}

To perform the PCA, we first divided the sample into training (70\,\%) and test (30\,\%) sets using stratified sampling to ensure that these sets are representative of the overall population (have the same percentage of samples from each target class as the complete set). Thus, we trained the PCA model using the training set, obtaining that 93\,\% of the sample's variance lied along the two first principal components. Projecting the training data onto the hyperplane defined by these two principal components, the vast majority of non-UCD objects are clearly separated from the UCDs. Thus, it is possible to make a first cut in the identification of UCDs with this 2D projection, by defining a decision threshold (purple line in the Figure) and keeping only the objects that fall on the UCD side. Fig. \ref{fig:pca} shows the same projection for the entire sample (training + test). After this cut, we reduced our sample to 317 and 29\,732 UCD and non-UCD objects, respectively, achieving a 94\,\% reduction on the negative class. Despite still being strongly imbalanced, this reduced sample has a better balance between the negative and positive class, which facilitates better results when using the SVM.

\subsection{SVM model} \label{svm}

To develop the SVM model, we used the reduced sample obtained in the PCA filtering, keeping the same training and test set structure. We used the test set for the validation of the classification model. The seven J-PLUS colours described in Sect. \ref{pca_cut} were used as features in the training step.

Then, we conducted a search for the SVM's optimal hyperparameters on the training test. To do this, we created a grid for the SVM kernel and hyperparameters and did an exhaustive search over this parameter space using the \texttt{GridSearchCV} class from the \texttt{scikit-learn} package, which optimises the hyperparameters of an estimator by k-fold cross-validation using any score to evaluate the performance of the model. In our case, we used the recall score, which measures the ability of the classifier to find all the positive instances, since our priority is to identify as many candidate UCDs as possible. For the \texttt{GridSearchCV} class, we used ten k-folds and set the hyperparameter \texttt{class\_weight} to `balanced' to address the imbalance by adjusting the weights inversely proportional to the class frequencies. In the grid of hyperparameters, we tested the regularisation parameter $C$ for values of 1, 10, 100 and 1000, and the kernel scale $\gamma$ of the RBF kernel for 0.001, 0.01, 0.1, 1, 10 and 100.

\begin{figure}
	\includegraphics[width=\columnwidth]{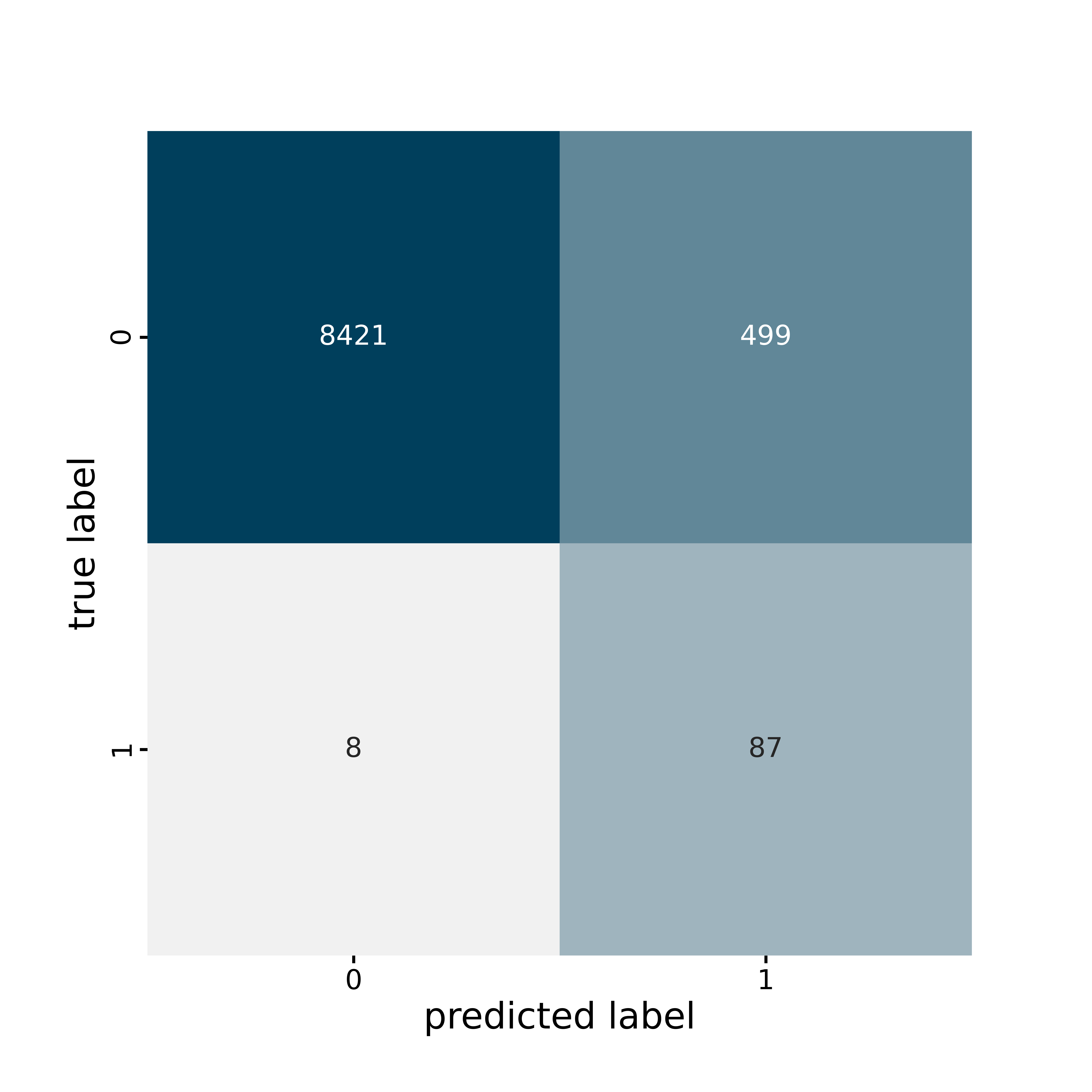}
    \caption{Confusion matrix for the test set, with a recall of 92\,\%.}
    \label{fig:conf_mat_tr}
\end{figure}  

After this search for the optimal hyperparameters, we obtained the best recall score with an RBF kernel and hyperparameters $C=10$ and $\gamma=0.001$, with a total recall of 92\,\% on the test set. Fig. \ref{fig:conf_mat_tr} shows the confusion matrix on the test set. The confusion matrix is a performance measurement in machine learning classification that compares the labels predicted by the model (x-axis) with the ground-truth labels in the data set (y-axis). The most important thing to note here is that the SVM model manages to recover nearly all positive instances, which is our main priority, as we do not want to lose any candidate UCD in the process. Also, the SVM performs very well at identifying True Negatives (TN, negative instances predicted as negative). In conclusion, the model allows us to filter out the vast majority of non-UCD objects, while keeping almost all the candidate UCDs. However, the class imbalance of the data causes the number of False Positives (FP, negative instances predicted as positive) to be larger than the number of True Positives (TP, positive instances predicted as positive). This makes the analysis with VOSA still necessary to differentiate the final candidate UCDs.

\subsection{Blind test} \label{blind_test}

\begin{figure}
	\includegraphics[width=\columnwidth]{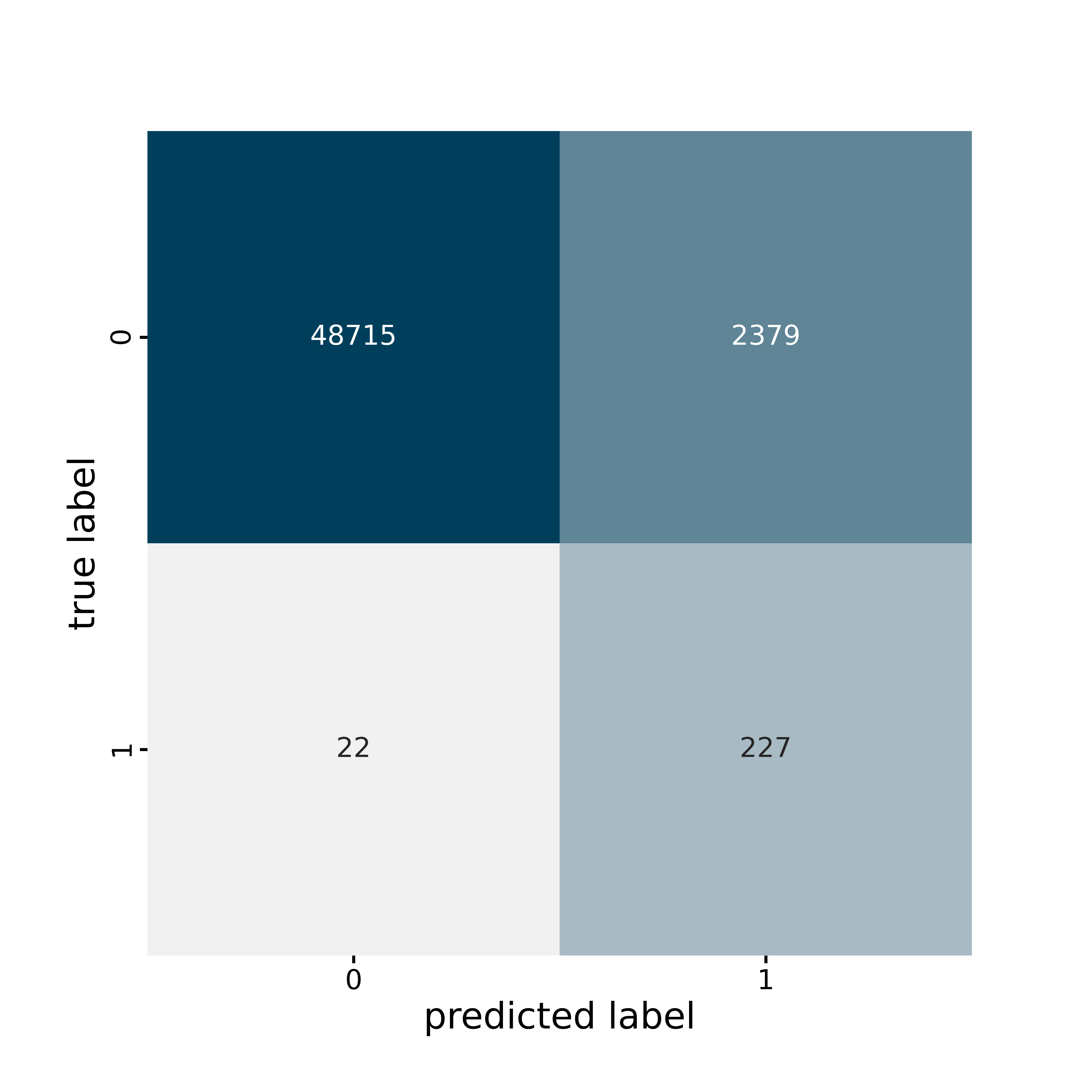}
    \caption{Confusion matrix for the blind test, with a recall of 91\,\%.}
    \label{fig:conf_mat_new}
\end{figure}

To validate the classifier's performance on unseen data, we applied our ML methodology on the J-PLUS DR2 data from another of the 20$\times$20\,deg$^2$ regions containing 607\,801 objects with good photometry in all relevant filters. Firstly, we used the same PCA model fitted with the previous region to perform the PCA filtering on this new region, reducing the total number of instances to 51\,343. We used the previously fitted SVM model to predict over this reduced set, obtaining a recall score of 91\,\%. Fig. \ref{fig:conf_mat_new} shows the confusion matrix for this blind test. Thus, we ended up with 2\,606 (2\,379 + 227) objects to be analysed with VOSA for the final UCD identification, which means the SVM model achieved to discard $\sim$95\,\% (1 - 2\,379/51\,094) of the non-UCD objects that pass the PCA filtering.

We used the objects analysed with \texttt{VOSA} in the VO methodology for this same region to make a thorough analysis of our ML method. Thus, we found that, of these objects, the PCA filter removes those with $T_{\rm eff}\gtrsim4\,100$\,K, so this first cut is able to purge the initial set of the hottest objects. The ML methodology is more restrictive in terms of photometric quality, as it is only applicable to the objects with photometry in all the filters used to build the input features. This means that all the final candidate UCDs with no photometry in any of these filters (around 50\,\% for this region), obtained with the VO methodology, are not captured by the ML procedure. In summary, we concluded that the ML methodology is more efficient in the sense that it allows for a greater number of true negatives (non-UCD objects) to be discarded prior to analysis with \texttt{VOSA}, although it is a more restrictive method as it relies only on the photometry of the J-PLUS filters used. Another advantage of the proposed ML approach is that it consists of a single process instead of the three separate ones required in the VO methodology.

%--------------------------------------------------------------------

\section{Detection of strong emission line emitters} \label{flares}

Strong emission lines have been detected serendipitously in UCD optical spectra, both as transient flaring  phenomena  \citep{Liebert1999,Liebert2003,Martin2001,Schmidt2007} as well as steady  features  \citep{Schneider1991,Mould1994,Martin1999,Burgasser2011}. Stellar flares, events powered by the sudden release of magnetic energy, that is converted to kinetic energy of electrons and ions due to magnetic reconnection in the stellar atmosphere, are a common phenomenon around M dwarfs. Works such as those presented in \citet{flare_xray}, \citet{Berger2010} and \citet{arecibo} have confirmed that optical, radio and X-ray flares do occur in UCDs. 

We decided to focus our search for strong emission on the H$\alpha$ and Ca~{\sc ii} H and K lines, important chromospheric activity indicators \citep{cincunegui}, which correspond to filters 11.0 ($J0660$) and 7.0 ($J0395$) in the J-PLUS filter system, respectively. Since this is a rare phenomenon, we decided to conduct this search on a larger sample of objects, including all the objects that met the $G - G_{\rm RP}$ and $r-z$ colour criteria presented in Sects. \ref{parallax}, \ref{pm} and \ref{colordiagram}. Therefore, since we did not apply the effective temperature cutoff, the search also covered spectral types hotter than those of the UCDs.

With this purpose, we developed a Python algorithm that detects any drop in magnitude in filters $J0395$ and $J0660$. Firstly, the algorithm joins the J-PLUS DR2 photometry obtained in the search described in Sect. \ref{Methodology} to the shortlisted objects obtained with the methodology described in Sect. \ref{parallax}, \ref{pm} and \ref{colordiagram}. Then, object by object, it computes the magnitude ratio between the filter of interest and all its neighbours. We chose as neighbours the filters 6.0, 8.0 and 9.0 for the filter 7.0 (Ca~{\sc ii} H and K) and the filters 1.0 and 3.0 for the 11.0 (H$\alpha$). If this ratio is lower than a fixed threshold value (entered by the user) for any neighbouring filter, the algorithm recognises a  possible strong line emitter and plots the photometry of the object. For the object to be recognisable, we need at least photometry in one of the neighbouring filters, so we can detect this emission peak. The algorithm receives as input a file with the candidate UCDs photometry and returns both the plotted photometry of the objects with possible strong emission and a table with the computed magnitude drop for each of them. We were permissive with the fixed threshold, so as not to discard any interesting object, and imposed a value of 0.96. Then, we visually inspected all the possible strong emitters detected by the algorithm given this threshold.

\begin{figure}
	\includegraphics[width=\columnwidth]{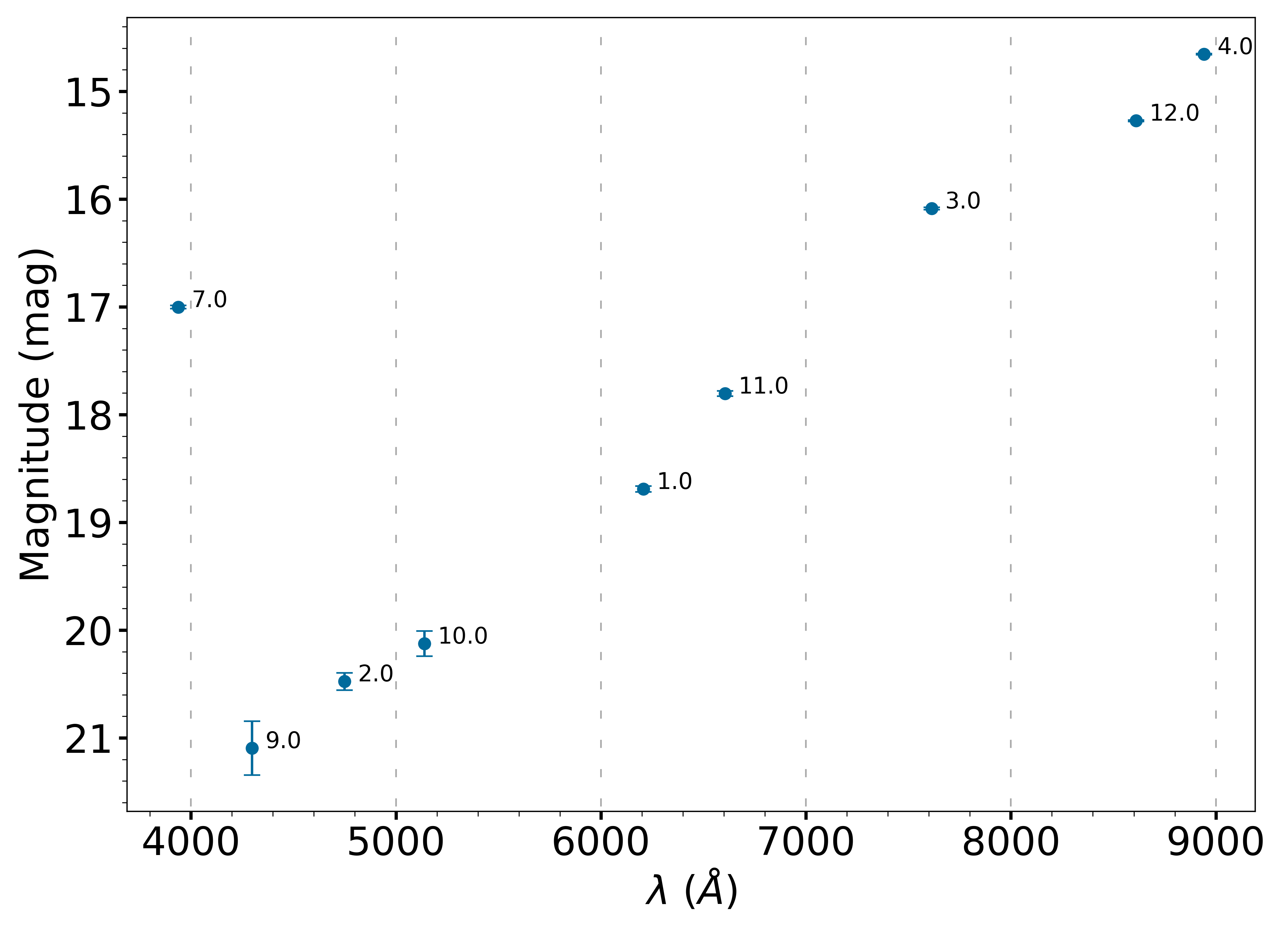}
    \caption{Example of a detected strong excess in the filter 7.0 ($J0395$), which corresponds to Ca~{\sc ii} H and K emission lines. The Figure shows the J-PLUS photometry of the first object in Table \ref{tab:flares}, with error bars representing the error in the magnitude. The algorithm detects that the magnitude in the filter $J0395$ is 0.8 times the magnitude in the filter 9.0 ($J0430$), and recognises this as a possible emission line. In this case, the threshold value for the excess detection was 0.96.}
    \label{fig:flares}
\end{figure}

\begin{figure*}
    \centering
    \includegraphics[width=\linewidth/6]{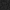}
    \hspace{0.1cm}
    \includegraphics[width=\linewidth/6]{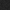}
    \hspace{0.1cm}
    \includegraphics[width=\linewidth/6]{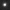}
    \hspace{0.1cm}
    \includegraphics[width=\linewidth/6]{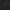}
    \hspace{0.1cm}
    \includegraphics[width=\linewidth/6]{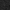}
    \\[\smallskipamount]
    \includegraphics[width=\linewidth/6]{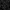}
    \hspace{0.1cm}
    \includegraphics[width=\linewidth/6]{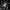}
    \hspace{0.1cm}
    \includegraphics[width=\linewidth/6]{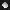}
    \hspace{0.1cm}
    \includegraphics[width=\linewidth/6]{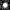}
    \hspace{0.1cm}
    \includegraphics[width=\linewidth/6]{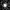}

    \caption{Images from the J-PLUS DR2 archive with the photometry in different filters for two of the strong line emitters detected. The first row corresponds to an excess in the Ca~{\sc ii} H and K (filter $J0395$) emission lines (first object in Table \ref{tab:flares}), with images in the filters $u$, $J0378$, $J0395$, $J0410$, and $J0430$ (from left to right). The second row shows an excess in the H$\alpha$ (filter $J0660$) emission line (seventh object in Table \ref{tab:flares}), with images in the filters $J0515$, $r$, $J0660$, $i$, and $J0861$ (from left to right). For both objects, all images shown were taken within a time interval of about 40 minutes.}
    \label{fig:flares_imgs}
\end{figure*}

Finally, we ended up with eight objects that exhibit significant emission peaks in the filters of interest, that are presented in Table \ref{tab:flares}. We used \texttt{VOSA} to estimate the effective temperature of these objects and found only one UCD, with $T_{\rm eff}=2\,500$\,K, among the eight objects (fifth object in Table \ref{tab:flares}). The remaining seven objects have estimated effective temperatures (see Table \ref{tab:flares}) typical of mid-M dwarfs \citep{zhang_2018_midm}. Fig. \ref{fig:flares} shows the photometry of the object with the highest line emission excess (first object in Table \ref{tab:flares}). Also, in Fig. \ref{fig:flares_imgs} we include images from the J-PLUS DR2 archive with the emission in different filters for the object with highest excess activity in the Ca~{\sc ii} H and K (first object in Table \ref{tab:flares}) and H$\alpha$ (seventh object in Table \ref{tab:flares}) emission lines. With this analysis, we underline the possibility of systematically detecting strong emission lines in UCDs and earlier M-type stars with photometric surveys such as J-PLUS.

For the fifth  object listed in Table \ref{tab:flares}, namely LP 310-34, we carried out a follow-up optical spectroscopy monitoring study. Five exposures of half an hour integration time each were obtained on January 12th, 2020 in service time (proposal 60-299, PI Martín) with ALFOSC attached to the Nordic Optical Telescope in La Palma. The grism number 4 and the slit with of 1.0 arcsec were selected providing a dispersion of 3.75 \AA pixel$^{-1}$ and a resolving power of R=700. Our spectra confirm that it is a very late M dwarf (dM8) with H$\alpha$ in emission \citep{Schmidt2007}. We measured an H$\alpha$ equivalent width of -14.6 \AA , using the gaussian profile integration option available in the IRAF task splot applied to the co-added spectrum of the five exposures. Individual measurements of the equivalent width in each spectrum ranged from -7.0 to -20.7 \AA , suggesting variability in the strength of the H$\alpha$ emission. This level of H$\alpha$ emission is not uncommon among late-M dwarfs \citep{Martin2010,Pineda2016}. No other emission lines were detected in our spectra. 

One of the new strong line-emission candidates (sixth object in Table \ref{tab:flares}) was observed on April 21st, 2022 with the long-slit low-resolution mode of the SpeX instrument \citep{Rayner2003} at the NASA Infrared Telescope Facility (IRTF, program 2022A011, PI A. Burgasser). Preliminary analysis of the data indicates that the near-infrared spectrum is well matched by a M5 dwarf template (A. Burgasser, private communication). Further details of these observations and additional spectroscopic follow-up of the J-PLUS candidates presented in this work is planned for a future paper.  

This study suggests that our J-PLUS search for strong emission lines may be revealing previously unknown sporadic very strong activity in otherwise normal late-M dwarfs. It is worth noting that our search for strong line emitters has detected as many objects with Ca~{\sc ii} H and K excess than with H$\alpha$ excess, and no object showing both excesses simultaneously. Events of strong Ca~{\sc ii} H and K line emission in normal late-M dwarfs may have important implications for studies of 
exoplanetary space weather and habitability \citep{Yamashiki2019}.

\begin{table*}
 \caption{Objects with strong flux excess in H$\alpha$ (filter $J0660$) or Ca~{\sc ii} H and K (filter $J0395$) emission lines, identified with our Python algorithm.}
 \label{tab:flares}
 \centering          
 \begin{tabular}{c c c c c c c}
  \hline\hline
  \noalign{\smallskip}
  
  $\alpha$ & $\delta$ & Filter of interest & Magnitude in filter of interest & Ratio of magnitudes & Simbad ID & Estimated $T_{\rm eff}$\\
  (deg) & (deg) &  &  & vs. neighbour filter  & & (K)\\
  
  \noalign{\smallskip}
  \hline
  \noalign{\smallskip}
  
  18.53140 & 7.94229 & $J0395$ & 16.811 & 0.799 & \ldots & 3\,200\\
  
  36.68415 & 34.75973 & $J0660$ & 18.100 & 0.884 & \ldots & 3\,300\\
  
  107.18550 & 71.90704 & $J0660$ & 18.199 & 0.915 & \ldots & 3\,200\\

  116.14374 & 40.14576 & $J0395$ & 19.333 & 0.946 & \ldots & 3\,100\\
  
  121.85651 & 32.21826 & $J0395$ & 17.002 & 0.806 & LP 310-34 & 2\,500\\

  135.92497 & 34.80495 & $J0395$ & 18.319 & 0.895 & LP 259-39 & 3\,200\\

  138.52385 & 23.87355 & $J0660$ & 17.443 & 0.850 & \ldots & 3\,200\\

  199.02058 & 56.12370 & $J0660$ & 20.233 & 0.911 & \ldots & 3\,300\\

  \noalign{\smallskip}
  \hline
 \end{tabular}
\end{table*}

%--------------------------------------------------------------------

\section{Conclusions and future work} \label{conclusions}

Using a Virtual Observatory methodology, we provide a catalogue of 9\,810 candidate UCDs over the entire sky coverage of J-PLUS DR2. With 7\,827 previously not reported as UCD, we show there is still room for the discovery of these objects even with a small telescope such as the JAST80. Our main goal is to consolidate and further develop a search methodology, introduced in \citet{solano2019}, to be used for deeper and larger surveys such as J-PAS and Euclid, both being an ideal scenario for the study and discovery of UCDs thanks to their unprecedented photometric system of 54 narrow-band filters and excellent sensitivity, respectively. Further confirmation by spectroscopy of the UCD nature of these candidates goes beyond the scope of this study. However, the candidate UCDs that are reported in Simbad, but are not in our sample of known UCDs (see Sect. \ref{known}), mostly present spectral type M6\,V or are left out because they lack the luminosity class, so we expect the degree of contamination to be small.

The use of different approaches based on astrometry and photometry tends to minimise the drawbacks and biases associated to the search of ultracool objects: photometric-only selected samples may leave out peculiar UCDs not following the canonical trend in colour-colour diagrams and they can also be affected by extragalactic contamination. Proper motion searches may ignore objects with small values of projected velocity in the plane of the sky. Regarding parallax-based searches, they will be limited to the brightest objects with parallax values from \textit{Gaia}.

Based on our kinematics study, almost all our candidate UCDs can be considered thin disk members, with 268 of them being potential members of the thick disk. Also, five of the candidates are likely to belong to the Galactic halo. Using the \texttt{BANYAN}~$\Sigma$ tool, we find 48 candidate UCDs with a high Bayesian probability of belonging to seven different young moving associations, in 30 of the cases with a probability greater than 95\,\%. A further spectroscopic follow-up will be required to search for spectral signatures of youth. In the binarity analysis, we find 122 possible unresolved companions among our candidate UCDs. Searching for wide \textit{Gaia} companions of our candidate UCDs, we find 78 possible multiple systems (73 binary + 5 triple), six of them already tabulated by the WDS. We use \texttt{VOSA} to get an estimation of the effective temperature of the wide \textit{Gaia} companions identified in all the systems, finding that most of them are M-type stars.

Among the non-recovered known UCDs that lie in the sky coverage of J-PLUS DR2, we find that more than half are lost due to lack of photometric or astrometric information with enough quality. The remaining objects are discarded due to our conservative temperature cutoff at 2\,900\,K or a bad SED fitting (vgfb$>$12). Compared to previously reported candidates, the new ones are on average more distant and extend to smaller values of proper motion.

We achieve promising results when reproducing the search for UCDs with a purely ML-based methodology. In this approach, we find crucial the preliminar PCA filtering to deal with the strong imbalance of the data and discard the hottest objects. This allows us to significantly reduce the negative class and improve the classification capability of the posterior SVM model. Using the developed ML methodology to predict on unseen data, we are able to recover 91\,\% of the candidate UCDs found with the VO methodology, discarding a larger number of true negatives (non-UCD objects) before the analysis with VOSA in a faster way. This is a significant achievement, since the main bottleneck of the VO methodology is the high number of objects to be analysed with \texttt{VOSA}.

In this line, the real turning point would be to develop a ML methodology that more significantly filters the number of objects we need to analyse with \texttt{VOSA} for the final UCD identification. This is not a straightforward task due to the imbalance of the data and because the analysis with \texttt{VOSA} is based on complex theoretical models. To this end, we are exploring the use of independent component analysis in the initial filtering and ensemble learning in the classification step.

Finally, we develop an algorithm capable of detecting strong emission line emitters in the optical range. We identify four objects with strong excess in the filter corresponding to the Ca~{\sc ii} H and K emission lines and four other objects with excess emission in the H$\alpha$ filter.

\begin{acknowledgements}

       This research has made use of the Spanish Virtual Observatory (https://svo.cab.inta-csic.es) project funded by MCIN/AEI/10.13039/501100011033/ through grant PID2020-112949GB-I00 and MDM-2017-0737 at Centro de Astrobiolog\'{i}a (CSIC-INTA), Unidad de Excelencia Mar\'{i}a de Maeztu. This work presents results from the European Space Agency (ESA) space mission \emph{Gaia}. \emph{Gaia} data are being processed by the \emph{Gaia} Data Processing and Analysis Consortium (DPAC). Funding for the DPAC is provided by national institutions, in particular the institutions participating in the Gaia MultiLateral Agreement (MLA). The \emph{Gaia} mission website is https://www.cosmos.esa.int/gaia. The \emph{Gaia} archive website is https://archives.esac.esa.int/gaia. This research has made extensive use of the SIMBAD database \citep{Wenger00}, VizieR catalogue access tool \citep{och2000}, Aladin sky atlas \citep{aladin} provided by CDS, Strasbourg, France, and of the TOPCAT \citep{Taylor2005} and STILTS \citep{Taylor2006} tools. We also made use of VOSA and the SVO Filter Profile Service, developed under the Spanish Virtual Observatory project. We made extensive use of Python throughout the entire process, including the packages \texttt{pandas}\footnote{\url{https://github.com/pandas-dev/pandas}}, \texttt{seaborn} \citep{seaborn}, \texttt{numpy} \citep{numpy}, \texttt{matplotlib} \citep{matplotlib}, \texttt{scikit-learn} \citep{sklearn}, \texttt{plotly}\footnote{\url{https://github.com/plotly/plotly.py}}, \texttt{scipy} \citep{scipy}, \texttt{astropy} \citep{astropy}, and \texttt{mocpy} \citep{mocpy}.
       EM was supported by the Agencia Estatal de Investigación del Ministerio de Ciencia e Innovación (AEI-MCINN) under grant PID2019-109522GB-C53. We want to thank Jonathan Gagné for the development of the BANYAN $\Sigma$ tool\footnote{\url{http://www.exoplanetes.umontreal.ca/banyan/}}. PMB is funded by INTA through grant PRE-OVE. F.J.E. acknowledges financial support by the Spanish grant MDM-2017-0737 at Centro de Astrobiolog\'{\i}a (CSIC-INTA), Unidad de Excelencia Mar\'{\i}a de Maeztu. Funding for the J-PLUS project has been provided by the Governments of Spain and Arag\'on through the Fondo de Inversiones de Teruel; the Aragonese Government through the Research Groups E96, E103, E16\_17R, and E16\_20R; the Spanish Ministry of Science, Innovation and Universities (MCIU/AEI/FEDER, UE) with grants PGC2018-097585-B-C21 and PGC2018-097585-B-C22; the Spanish Ministry of Economy and Competitiveness (MINECO/FEDER, UE) under AYA2015-66211-C2-1-P, AYA2015-66211-C2-2, AYA2012-30789, and ICTS-2009-14; and European FEDER funding (FCDD10-4E-867, FCDD13-4E-2685). The Brazilian agencies FAPERJ and FAPESP as well as the National Observatory of Brazil have also contributed to this project. This work is based on observations made with the JAST80 telescope at the Observatorio Astrofísico de Javalambre (OAJ) in Teruel, owned, managed and operated by the Centro de Estudios de Física del Cosmos de Aragón (CEFCA) and data obtained with ALFOSC on the Nordic Optical Telescope (NOT) at the Observatorio del Roque de los Muchachos (La Palma, Spain). ALFOSC was provided by the Instituto de Astrofísica de Andalucía (IAA) under joint agreement with the University of Copenhaguen and NOTSA. We thank the OAJ Data Processing and Archiving Unit (UPAD) for reducing and calibrating the OAJ data used in this work. R.A.D. acknowledges support from the Conselho Nacional de Desenvolvimento Científico e Tecnológico – CNPq through BP grant 308105/2018-4, and the Financiadora de Estudos e Projetos – FINEP grants REF. 1217/13 – 01.13.0279.00 and REF 0859/10 – 01.10.0663.00 for hardware support for the J-PLUS project through the National Observatory of Brazil. PC acknowledges financial support from the Government of Comunidad Autónoma de Madrid (Spain), via postdoctoral grant ‘Atracción de Talento Investigador’ 2019-T2/TIC-14760. We thank the IRTF observers (D. Chih-Chun Hsu, N. Lodieu, C. Theissen, J.-Y. Zhang) for program 2022-A011 (PI A. Burgasser). The authors wish to recognise and acknowledge the very significant cultural role and reverence that the summit of Maunakea has always had within the indigenous Hawaiian community.  We are most fortunate to have the opportunity to conduct observations from this mountain.

\end{acknowledgements}

\bibliographystyle{aa} % style aa.bst
\bibliography{Bibliography} % your references Yourfile.bib

\appendix

\section{Virtual Observatory compliant, online catalogue} \label{vo_catalogue}

To help the astronomical community use our catalogue of candidate UCDs, we provide an archive system that can be accessed  from  a webpage\footnote{\url{http://svocats.cab.inta-csic.es/jplus_ucds1; http://svocats.cab.inta-csic.es/jplus_ucds2}} or through a Virtual Observatory ConeSearch\footnote{e.g.\url{http://svocats.cab.inta-csic.es/jplus_ucds1/cs.php?RA=0.023&DEC=35.457&SR=0.1&VERB=2; http://svocats.cab.inta-csic.es/jplus_ucds2/cs.php?RA=238.569&DEC=52.742&SR=0.1&VERB=2}}. The archive system implements a  very simple search interface that allows queries by coordinates and radius as well as by other parameters of interest. The user can also select the maximum number of sources (with values from ten to unlimited). The result of the query is a HTML table with all the sources found in the archive fulfilling the search criteria. The result can also be downloaded as a VOTable or a CSV file. Detailed information on the output fields can be obtained placing the mouse over the question mark located close to the name of the column. The archive also implements the SAMP\footnote{\tt http://www.ivoa.net/documents/SAMP} (Simple Application Messaging)  Virtual  Observatory  protocol. SAMP allows Virtual Observatory  applications to communicate with each other in a seamless  and transparent manner for the user. This way, the results of a query can be easily transferred  to  other  VO  applications, such as, for instance, \texttt{TOPCAT}.

\end{document}